\documentclass{aa}
\usepackage[varg]{txfonts}
\pdfoutput=1
\usepackage{graphicx}
\usepackage{newtxtext,newtxmath}

\usepackage{textcomp}
\usepackage[inline]{enumitem}
\usepackage{amsmath}
\usepackage{hyperref}
\usepackage{float}
\usepackage{placeins}
\usepackage{natbib}
\usepackage{amsmath, amssymb, amsfonts}
\usepackage{cmdDef}
\usepackage{xcolor}
\usepackage[normalem]{ulem} 

\newcommand\hl{\bgroup\markoverwith
  {\textcolor{yellow}{\rule[-.5ex]{2pt}{2.5ex}}}\ULon}

\newcommand{\msun}{\,M$_{\odot}$\xspace}

\begin{document}

    \title{The effects of environment on galaxies' dynamical structures:\\ From simulations to observations}    
    \titlerunning{Environmental effects on galaxies' dynamical structures}
    \authorrunning{Y. Ding et al.}

    \author{%
        Y.~Ding,\inst{1,2}\thanks{Email:ycding@shao.ac.cn},
        L.~Zhu\inst{1}\thanks{Corr author: lzhu@shao.ac.cn},
        A.~Pillepich\inst{3},
        G.~van de Ven\inst{4},
        E.~M.~Corsini\inst{5,6},
        E.~Iodice\inst{7},
        F.~Pinna\inst{8,9,10}
}
    \authorrunning{Y. Ding et al.}

    \institute{%
        Shanghai Astronomical Observatory, Chinese Academy of Sciences, 80 Nandan Road, Shanghai 200030, China\and
        School of Astronomy and Space Sciences, University of Chinese Academy of Sciences, No. 19A Yuquan Road, Beijing 100049, China\and
        Max-Planck-Institut für Astronomie, Königstuhl 17, D-69117 Heidelberg, Germany\and
        Department of Astrophysics, University of Vienna, T{\"u}rkenschanzstra{\ss}e 17, 1180 Wien, Austria\and
        Dipartimento di Fisica e Astronomia “G. Galilei”, Università di Padova, Padova, Italy\and
        INAF - Osservatorio Astronomico di Padova, Padova, Italy\and
        INAF - Astronomical Observatory of Capodimonte, Salita Moiariello 16, 80131, Naples, Italy\and
        Max Planck Institute for Astronomy, Koenigstuhl 17, D-69117 Heidelberg, Germany\and
        Instituto de Astrofísica de Canarias, Calle Vía Láctea s/n, E-38205 La Laguna, Tenerife, Spain\and
        Departamento de Astrofísica, Universidad de La Laguna, Av. del Astrofísico Francisco Sánchez s/n, E-38206, La Laguna, Tenerife, Spain
    }

    \date{Received XXXXX; accepted XXXXX}

    \abstract{
We studied the effects of cluster environments on galactic structures by using the TNG50 cosmological simulation and observed galaxies in the Fornax cluster. We focused on galaxies with stellar masses of $10^{8-12}$\msun at $z=0$ that reside in Fornax-like clusters with total masses of $M_{200c} = 10^{13.4-14.3}$\msun.
We characterized the stellar structures by decomposing each galaxy into a dynamically cold disk and a hot non-disk component, and studied the evolution of both the stellar and gaseous constituents.
In TNG50, we find that the cold (i.e., star-forming) gas is quickly removed when a galaxy falls into a Fornax-mass cluster. About 42\%, 73\%, and 87\% of the galaxies have lost $80\%$ of their star-forming gas at 1, 2, and 4 billion years after infall, respectively, with the remaining gas concentrating in the inner regions of the galaxy. 
The radius of the star-forming gaseous disk decreases to half its original size at 1, 2, and 4 billion years after infall for 7\%, 27\%, and 66\% of the galaxies, respectively.
As a result, star formation (SF) in the extended dynamically cold disk sharply decreases, even though a low level of SF persists at the center for a few additional gigayears. This leads to a tight correlation between the average stellar age in the dynamically cold disk and the infall time of galaxies. Furthermore, the luminosity fraction of the dynamically cold disk in ancient infallers (i.e., with an infall time $\gtrsim$ 8 Gyr ago) is only about one-third of that in recent infallers (infall time $\lesssim$ 4 Gyr ago), controlling for galaxy stellar mass. This quantitatively agrees with what is observed in early-type galaxies in the Fornax cluster. Gas removal stops the possible growth of the disk, with gas removed earlier in galaxies that fell in earlier, and hence the cold-disk fraction is correlated with the infall time. The stellar disk can be significantly disrupted by tidal forces after infall, through a long-term process that enhances the difference among cluster galaxies with different infall times.
    }


    \maketitle

\section{Introduction}
\label{sec:intro}

The cosmological environment of galaxies is thought to play an important role in their evolution. In general, observations have shown that galaxies in dense environments, such as those in galaxy groups and clusters, have a higher probability of being elliptical, red, and quenched (early-type galaxies), whereas galaxies ``in the field'' are more likely to be spiral, blue, and star-forming \citep[late-type galaxies; e.g.,][]{dressler1980,Butcher1984,Dressler1997,Lewis2002,blanton2005,alpaslan2015}. 

Three main physical processes are commonly invoked to explain this systematic phenomenology, all of which are related to gas removal: ``ram pressure'' exerted by the intracluster medium \citep{Gunn1972,Abadi1999,Yun2019}, ``harassment'' by fly-by galaxies \citep{Gunn1972,toomre1972,Barnes1992,Bournaud2004}, and ``strangulation'' caused by the cutoff of gas accretion \citep{Larson1980,Balogh2000,Kawata2008}. The dependence of environmental effects on the mass of the host halo and on the galaxy's stellar mass has been well characterized, in both large and more targeted galaxy surveys \citep[e.g.,][]{peng2010, Li2020ApJ...902...75L}. 
 
However, the details of environmental quenching, namely the stopping of star formation (SF) in group and cluster galaxies, are still debated. For example, the reduction in or halting of SF as a result of gas removal has typically been assumed to be a fast process \citep[e.g.,][]{baxter2023}. On the other hand, observations of galaxies in the Hydra cluster seem to indicate that it may take more than 600 Myr after infall to significantly remove their gas \citep{Wang2021}. Studies of low-redshift and local clusters also show that a smooth decrease in the star formation rates (SFRs) of satellite galaxies can be a continuous process lasting 2-5 Gyr, followed by a more rapid 1 Gyr quenching phase \citep{wetzel2013, vulcani2018,finn2023}. By comparing the output of the Millennium simulation (equipped with a semi-analytical model for galaxy formation) and observations of the Local Cluster Substructure Survey (LoCuSS), it has been suggested that SF in cluster galaxies can last for 1-2 Gyr after their infall into the cluster and then quickly gets quenched, a scenario dubbed ``slow-then-rapid quenching'' \citep{Maier2019}. Cosmological hydrodynamical galaxy simulations have shown that, for satellites in Milky Way-like halos for example, the quenching timescale is anticorrelated with the impulsiveness of the ram pressure, with a stronger impulsiveness of ram pressure implying shorter quenching timescales, and vice versa \citep[][with FIRE-2]{samuel2023}. Recently, with the IllustrisTNG simulations, \cite{rohr2023} have shown that, across a wide range of host and satellite masses, ram-pressure stripping (RPS) may act for long periods of time (up to several billion years), but with peak RPS periods of 1-2 Gyr after infall, such that SF continues until about 98 percent of the cold gas is lost. The situation is further complicated by the fact that SF may even be enhanced within 500 Myr after the onset of ram pressure \citep[e.g.,][]{kronberger2008,kapferer2009}, as also shown in simulations \citep{goeller2023}.

Importantly, the quenching of SF is thought to affect the makeup of galaxies' internal structures, adding to the effects of long-term dynamical processes after cluster infall that are also expected to affect the evolution of galaxies' stellar and gaseous morphology and content. Observations in the last decade have illuminated these connections. Firstly, integral-field spectroscopic (IFS) observations have provided stellar kinematic maps for thousands of nearby galaxies, showing that the red quenched early-type galaxies (ETGs) in fact exhibit large variations in internal kinematic properties, from slow- to fast-rotators \citep{Cappellari2011,weijmans2014}.

It has also been determined that, in the nearby Fornax, Virgo, and Coma clusters, slowly rotating ETGs are mainly located toward the cluster centers or in local overdense regions \citep{cappellari2011a, scott2014, cappellari2013b} and are thus likely ancient infallers \citep{Rhee2017}, whereas fast-rotating ETGs tend to populate the outskirts and are thus likely more recent infallers.  
By performing a dynamical decomposition for ETGs with high-quality integral-field unit (IFU) data from the Multi-Unit Spectroscopic Explorer (MUSE) as part of the Fornax3D project \citep{sarzi2018}, in previous work \citep{Ding2023} we quantified the stellar structure of galaxies by the luminosity fraction of their dynamically cold disks. For these Fornax ETGs, the cold-disk fraction in ancient infallers (i.e., satellites that fell into the cluster more than 8 Gyr ago) is only about one-third that of more recent infallers, controlling for galaxies' stellar mass. However, from observations alone we cannot directly tell how much of this difference is caused by preprocessing \citep[i.e., galaxy properties being affected by other high-density environments prior to infall into the $z=0$ cluster; e.g.,][]{donnari2021a}, by population biases (e.g., stellar structures of future satellites already differing at infall), or by physical processes affecting galaxies after they fall into their clusters, including the halting of SF. To get insights into this complex interplay of mechanisms, theoretical models of galaxies and clusters are of the essence.

Modern cosmological hydrodynamical simulations of galaxies allow detailed analyses of how different physical processes affect the stellar and gaseous structures of galaxies across cosmological environments. In particular, the IllustrisTNG suite \citep{pillepich2018b, nelson2018, naiman2018, marinacci2018, springel2018, nelson2019p} provides tens of thousands of well-resolved galaxies across three orders of magnitude of host halo masses, from Milky Way-like groups \citep{engler2021} to the massive galaxy clusters \citep{rigg2022}. The three flagship simulations, TNG100, TNG300, and the highest-resolution TNG50 \citep{pillepich2019,nelson2019}, have been shown to return populations of group and cluster satellites whose SFRs, quenched fractions, and atomic and molecular gas content are, to the first approximation, well in the ball park of observational constraints \citep[e.g.,][]{stevens2019, joshi2021, donnari2021b, stevens2021, engler2023}.

In terms of gas content and removal, thousands of jellyfish galaxies (i.e., galaxies exhibiting clear signatures of ongoing RPS) have been identified in the TNG100 and TNG50 simulations at $z\lesssim1$ \citep{Yun2019, zinger2023}, allowing connections to be drawn with observations and the aforementioned insights on the timescales of cold gas mass loss, SF, and SF quenching \citep{rohr2023, goeller2023}. Recently, it has been shown that TNG50 galaxies in high-density environments reproduce, albeit only qualitatively, the trends of gas truncation and central density suppression seen in the VERTICO survey in both HI and H$_2$ \citep{stevens2023} and that TNG100 satellites exhibit higher levels of HI asymmetries than central galaxies, due to environmental effects \citep{Watts2020}.

In terms of stellar structures, \cite{Joshi2020} demonstrated for the first time with fully cosmological hydrodynamical models (i.e., with TNG100 and TNG50) that cluster galaxies undergo morphological transformation, a notion that had until then been tested and evaluated only through a landscape of controlled non-cosmological and/or N-body-only numerical experiments. Tidal shocking was found to significantly disrupt the stellar disks of satellite galaxies within about $\sim 0.5-4$ Gyr after falling into Virgo-like clusters. According to the IllustrisTNG simulations, galaxies falling into the cluster earlier are more affected. Similar conclusions have been obtained for satellites in Fornax-mass clusters \citet{GalandeAnta2022}. Two physical processes were identified to justify the effects on the structures, and especially the survival of stellar disks, in satellites: (1) gas removal, chiefly via RPS (in combination with gas ejection via stellar and supermassive black hole feedback), which reduces the SF and the possible stellar disk growth, and (2) disruption of the existing stellar disk by tidal shocking at pericenter, rather than tidal stripping of the outer disk stellar material.

These results, based on the outcome and analysis of the IllustrisTNG simulations, qualitatively agree with the observational findings, that the surviving disk fraction in cluster galaxies is correlated with their infall times \citep[see above and][]{Ding2023}. However, previous studies have characterized galaxy structures and stellar disks from a theoretical viewpoint only, without the possibility to quantitatively compare such characterizations to observational results. In this paper, we build upon the previous IllustrisTNG findings and bridge the gap between simulation- and IFU-based characterizations of cluster galaxies.

We used the outcome of the TNG50 simulation and extracted galaxy stellar structures exactly in the same way as done for real galaxies in the Fornax cluster \citep{Ding2023}, in order to compare the simulation outcomes and the observations more directly. We selected simulated galaxies at $z=0$ in the observed stellar mass ranges that live in simulated clusters with masses comparable to that of the Fornax cluster. We followed the simulated galaxies back in time and quantified their gas content, SFRs, stellar age distributions, and the connection between the cold-disk fractions and infall times. 

The rest of the paper is organized as follows. We introduce the data and all operation definitions in Sect.~\ref{sec2}. We present the details of the environmental effects on the dynamically cold disk in Sect.~\ref{sec3}. We directly compare the real Fornax cluster and a TNG50 Fornax-like cluster in Sect.~\ref{sec4}, and we summarize our findings in Sect.~\ref{sec:conclusions}.

\section{Observations and the simulation}\label{sec2}
In this section we define the dynamical structures of TNG50 galaxies, extracted in ways that are comparable with what done observationally in the Fornax cluster, and define some key quantities used in the characterization of the results from the simulation.

\subsection{Observations of the Fornax cluster and its galaxies}
The Fornax cluster has a virial radius of $R_{\mathrm{vir}} \sim 0.7$ Mpc, a virial mass of $M_{\mathrm{vir}} \sim 7\times10^{13}$ M$_{\odot}$, and is located at a distance of $D \sim 20$ Mpc \citep{Diaferio1999,drinkwater2001}. Fornax and (most of) its galaxies have been observed with deep photometric images with the Fornax Deep Survey \citep[FDS;][]{venhola2018} and with high quality IFU data by Multi Unit Spectroscopic Explorer/Very Large Telescope (MUSE/VLT) through the Fornax3D project \citep{sarzi2018}.

In our previous work \citep{Ding2023}, we applied a population-orbit superposition method to 16 galaxies in the context of the Fornax3D project \citep{sarzi2018}. By fitting the luminosity distribution, stellar kinematics, age, and metallicity maps simultaneously, we obtained the internal stellar orbit distribution, as well as the age and metallicity distributions of stars on different orbits for each galaxy. 
Based on the model, we have been able to decompose each galaxy into dynamically cold disk and hot non-disk components, and define the luminosity fraction and stellar age distribution of each component. 
As mentioned in the Introduction, the cold-disk fraction in ancient infallers (fell into the cluster more than 8 Gyr ago) is only about one-third of the recent infallers, controlling for galaxies' stellar mass for ETGs in the Fornax cluster.
Such a structural decomposition can be done for simulated galaxies in exactly the same way, as we introduce next.

\subsection{The TNG50 simulation}
In this paper we use the public data \citep{nelson2019p} of TNG50 \citep{pillepich2019,nelson2019}, a cosmological magneto-hydrodynamical simulation covering a volume of $(35h^{-1}\mathrm{Mpc})^3$. TNG50 assumes a flat $\Lambda$ cold dark matter ($\Lambda$CDM) cosmology, with cosmological parameters from \citet{Collaboration2014}: density parameters $\Omega_m=0.3089$, $\Omega_{\Lambda}=0.6911$, $\Omega_b=0.0486$, normalization $\sigma_8=0.8159$ and spectral index $n_s=0.9667$. It was run with the moving-mesh code AREPO \citep{springel2010} and the IllustrisTNG galaxy formation model \citep{weinberger2017,pillepich2018} from $z=127$ to the current epoch. The data were saved at 100 snapshots from $z \sim 20$ to $z=0$.

The TNG50 simulation is initialized with a constant number of $2160^3$ dark matter particles and an initial number of $2160^3$ gas cell in its cubic comoving volume, resulting in a constant mass $ m_{\mathrm{DM}}=4.5\times10^5$\msun for dark matter particles and an average mass of $m_\mathrm{baryon}=8.5\times10^4$\msun for gas cells and stellar particles. The gravitational softening length is fixed at 288 pc from $z=0$ to $z=1$ for dark matter and stellar particles, whereas it is adaptive for the gas: it is set to 2.5 times the comoving gas cell radius, but with a minimum at 73.8 pc \citep[see][for details]{pillepich2019}.

TNG50 reproduces a large number of galaxies -- many thousands with stellar mass above $10^8$\msun -- with well-resolved stellar and gaseous structures that statistically match extragalactic observations in many contexts, including in the mass-size relation across cosmic epochs \citep{pillepich2019, Zanisi2021, Varma2022}, gas kinematics \citep{pillepich2019, ubler2021}, properties of, for example, stellar bars \citep{frankel2022}, fine stellar morphological structures \citep{Zanisi2021}, and internal dynamical structures as a function of stellar mass \citep{Xu2019,le2023}.

\subsubsection{Galaxies and galaxy clusters in TNG50}
In the simulated volume, halos (i.e., galaxy groups) are identified by a friends-of-friends \citep[FoF;][]{Davis1985} algorithm. 
The FoF algorithm is run on the dark matter particles, and the baryonic content in each FoF is assigned to the closest bound dark matter particle. 
The virial radius of a group is represented by $R_{200c}$ where the mean density is 200 times the critical density of a flat universe. Similarly, the virial mass $M_{200c}$ is the total mass enclosed in the virial radius $R_{200c}$. In this work, we selected all FoF groups with $M_{200c} = 10^{13.4-14.3}$ \msun at $z=0$, which have IDs from 0 to 13 in the catalog at the snapshot 099: they constitute our sample of simulated galaxy clusters, 14 in total.
In Sect.~\ref{sec4} we take Group ID 2 as a Fornax analog, as its virial mass and radius are the closest to those of the Fornax cluster.

Galaxies and subhalos are identified by the SUBFIND algorithm \citep{Springel2001}, and each subhalo with nonzero stellar mass is taken as a galaxy.
In each of the 14 clusters, we selected galaxies that are within the FoF group and located within the virial radius $R_{\mathrm{200c}}$ of their host halo at $z=0$. We only included galaxies with stellar mass $M_{*} = 10^{8 - 12}$\msun, measured as the total stellar mass of all stellar particles that are bound to the galaxy with the SUBFIND algorithm and with 2D effective radius $R_{\mathrm{e}}>0.5$ kpc. To compare directly with the observations, we used the r-band light-weighted 2D effective radius, $R_{\mathrm{e}}$, calculated by randomly projecting each galaxy on a 2D plane \citep[see][for details]{Zhu2022a}.

We also excluded the central galaxy of each cluster in order to exclusively focus on the environmental effects on the cluster satellite galaxies.

Finally, a fraction of cluster satellites have in fact been preprocessed \citep[e.g.,][]{donnari2021a}, meaning they had previously fallen into another group or cluster different from their $z=0$ host cluster, and then fallen into the host cluster via a cluster--cluster merger. For our analysis, we excluded all such preprocessed galaxies by searching the merger trees and checking if they ever fell into any other groups or clusters different from their $z=0$ host. Our final TNG50 galaxy sample includes 490 satellite galaxies in 14 clusters.

\begin{figure}
    \centerline{
        \includegraphics[width=0.85\columnwidth]{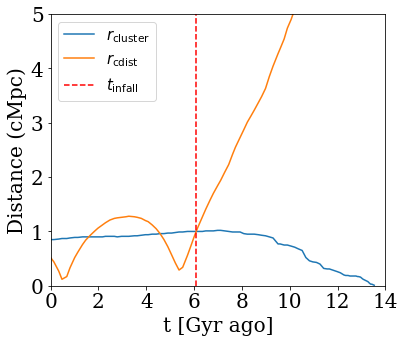}}
    \caption{{Visualization of the definition of infall time.} We show the cluster-centric distance of an example galaxy from TNG50 (SUBFIND ID 96768, orange curve) and the virial radius, $r_{200c}$, of its host halo (blue curve) as a function of the lookback time in gigayears. The vertical dashed line marks when the galaxy crosses the virial radius of the host cluster for the first time, which we define as its infall time, $t_{\rm infall}$. 
    }
    \label{def_infall}
\end{figure}
\subsubsection{Definition of infall time}
\label{SS:tinfall}

Though we selected cluster satellite galaxies at $z=0$, a fundamental notion to consider is for how long such galaxies have experienced the high-density environment of their host. For this, we need a time of infall.

For each galaxy in a cluster selected at $z=0$, we traced back its moving orbit and compared it to the progenitor of the $z=0$ host cluster via the merger trees.
Doing so, we found the time it crosses the virial radius of the $z=0$ host cluster for the first time; we defined this as the galaxy's infall time into the cluster, $t_{\rm infall}$. 
 
As an example, in Fig.~\ref{def_infall}, we show the 3D cluster-centric distance of the $z=0$ galaxy with SUBFIND ID 96768 as a function of cosmic time. By comparing to the 3D virial radius of its host cluster, we can see that this galaxy crossed the cluster virial radius for the first time $\sim 6$ Gyr ago. We thereby defined the infall time of this galaxy as $t_{\rm infall} = 6$ Gyr ago. We define ancient, intermediate, and recent infallers as those galaxies that fell into their host cluster 8-12,  4-8, and 0-4 Gyr ago, respectively.

\subsubsection{Gas and SF properties}
\label{SS:gasprops}

For the purposes of this work, we focused exclusively on the ``cold gas'' content of galaxies. By cold gas, we mean star-forming gas. The cold gas mass of each simulated galaxy is measured by summing up the mass of all gas cells with SFR $>0$ and that are bound to the galaxy according to the SUBFIND algorithm. We adopted the instantaneous SFR of each gas cell as output by the simulation. In the following sections, by ``gas'' we always mean cold (i.e., star-forming) gas, even if the characterization is omitted for brevity.

\subsection{Galaxy structure decomposition} 
\label{SS:decomposition}
In this work, we focused on dynamically defined stellar components of galaxies. Namely, we dynamically decomposed a galaxy into different structures based on their stellar orbit distribution.

In previous observational work, we obtained a galaxy's stellar orbit distribution from a population-orbit superposition method and described it as the probability density distribution in the phase-space of circularity $\lambda_z$ versus. radius $r$. The circularity $\lambda_z=J_z/J_{\mathrm{max}}(E_b)$ is the angular momentum $J_z$ aligned with the short axis $z$ normalized by the maximum angular momentum that is allowed by a circular orbit with the same binding energy $J_{\mathrm{max}}(E_b)$. The radius $r$ is the averaged radius of stars sampled along the orbit with equal time steps.
With the simulation data at hand, the orbital information is not directly available in that the data have been stored only every 100-200 Myr of temporal resolution \citep{nelson2019p} and a reconstruction and integration of the orbits would need to be done in post-processing. Instead, we have the 6D instantaneous phase-space information of each stellar particle. To make the simulation and observation comparable, we used a phase-space-average method to approximately obtain $\lambda_z$ and $r$ of the orbits of particles in simulated galaxies, following \citet{zhu2018a}, which also showed the level of consistency between the two approaches (see their appendix).

We first separated all the particles into bins in a 3D space of angular momentum $J_z$, total angular momentum $J$, and binding energy $E_b$. Assuming that particles in the same bin share similar orbits and thus similar orbital quantities, we took the average value of the instantaneous $\lambda_z$ and $r$ of particles in each bin as proxies of the true orbital $\lambda_z$ and $r$ of particles in that bin. 
We thus obtained the stellar orbital distribution for each TNG50 galaxy by weighting the stellar particles by their r-band luminosity, which is similar to that of real galaxies. This luminosity was estimated from the r-band Vega magnitude of each stellar particle assuming a Bruzual \& Charlot single-age stellar population model.

\begin{figure}
    \centerline{
        \includegraphics[width=1.\columnwidth]{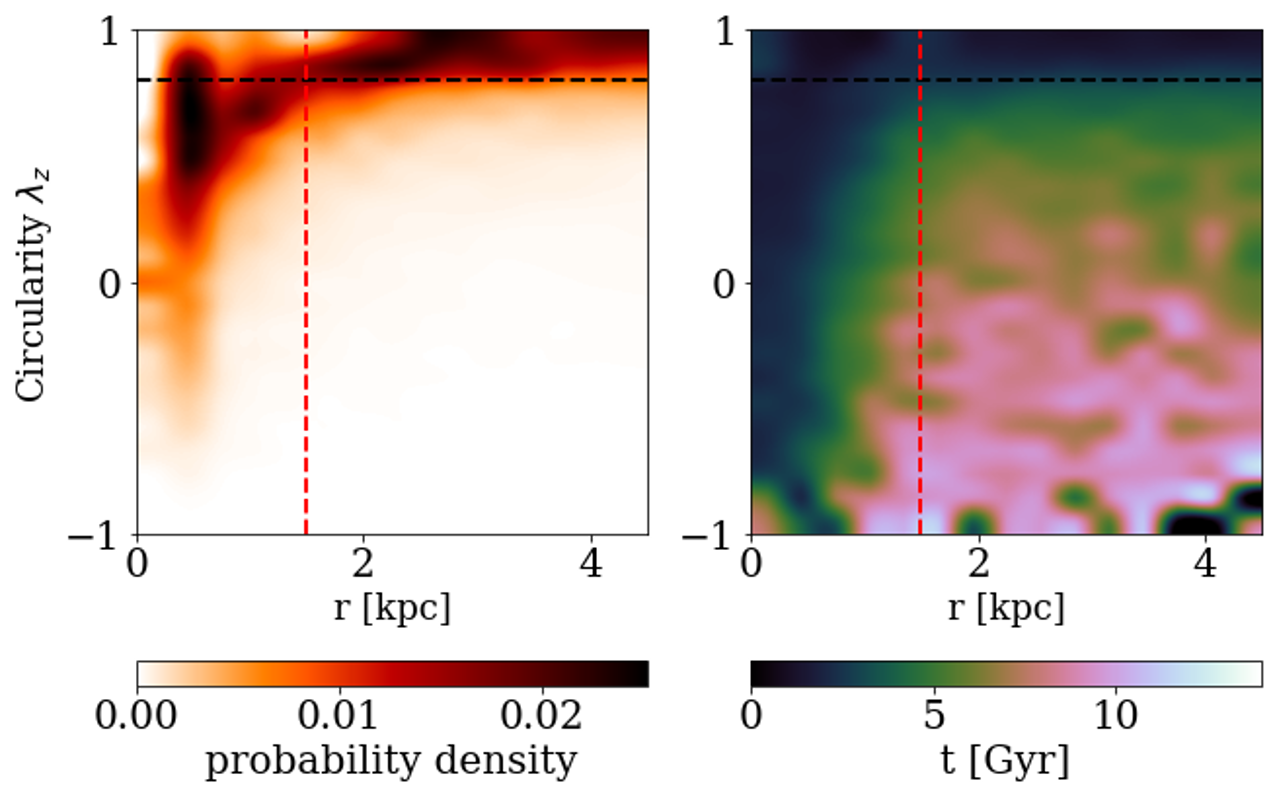}}
    \caption{
    {Orbital decomposition of a typical TNG50 galaxy.} We show the probability density
    distribution, $p(r, \lambda_z)$ {\em (left panel)\/}, and age distribution,
    $p(r, t)$ {\em (right panel)\/} of the stellar orbits in the phase space of
    radius $r$ versus circularity $\lambda_z$. The
    probability densities are normalized to unity within $3R_{\mathrm{e}}$. The dashed black line marks our division into two stellar components: a dynamically cold disk component ($\lambda_z \geq 0.8$) and a dynamically hot non-disk component ($\lambda_z <
    0.8$). The dashed red line represents the effective radius, $R_{\rm e}$, of this galaxy.}
    \label{orbital_decmp}
\end{figure}

\begin{figure*}
    \centerline{
        \includegraphics[width=1.7\columnwidth]{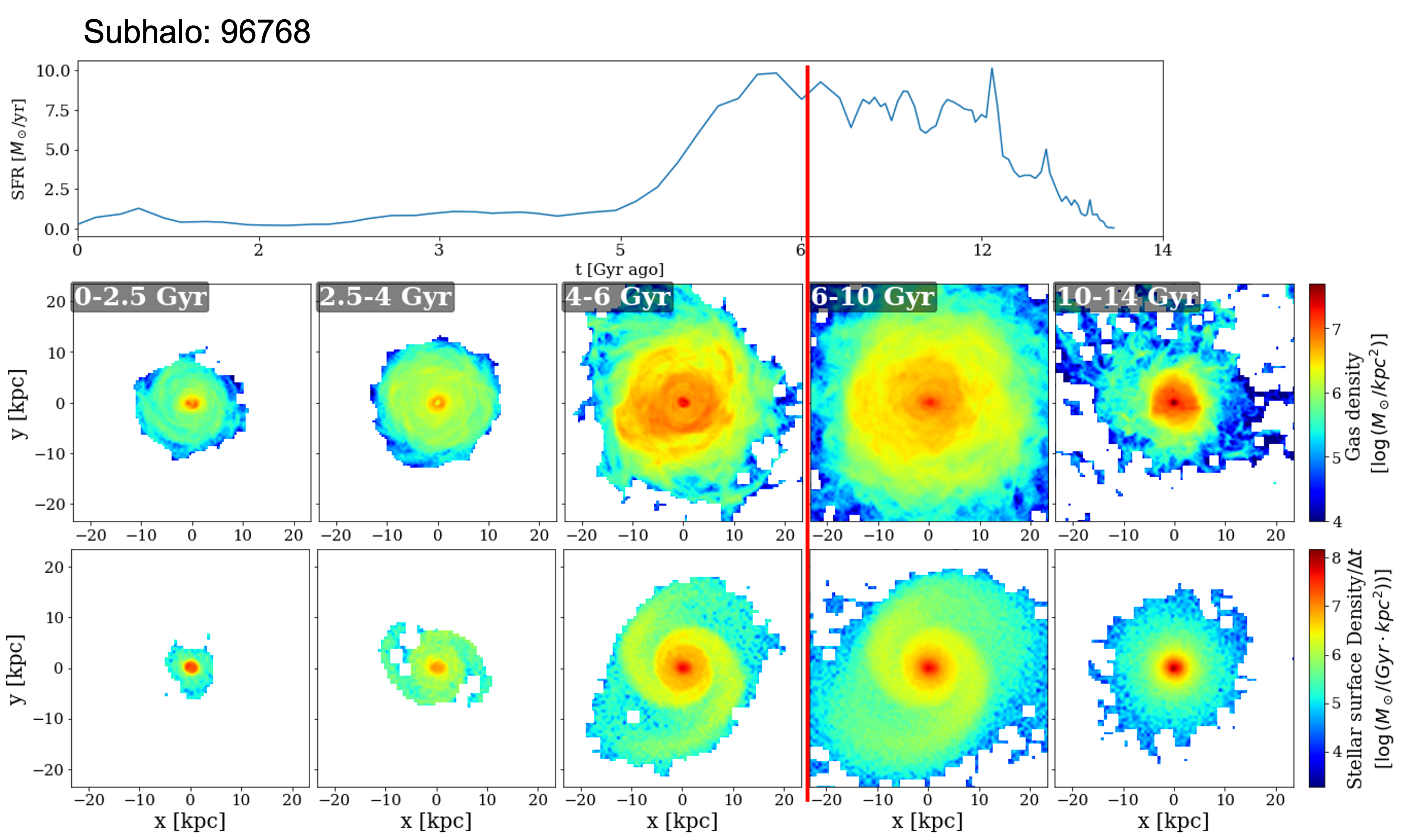}}
    \caption{{Evolution of an example cluster galaxy from TNG50 (Subhalo ID 96768 at $z=0$).} We show the time evolution of the SFR {\em (top panel)\/}, the average star-forming gas density {\em (middle panels),\/} and a face-on view of the average stellar surface density {\em (bottom panels)\/} of this galaxy that at $z=0$ has $10^{10.5}$\msun in stars and is a satellite of a cluster of $10^{13.8}$\msun in total mass. For the average stellar surface density, we do not show the property of the galaxy at the various epochs, but rather the stellar distribution in different age bins, from left to right and normalized by the length of age interval in each bin; the maps are obtained from all 
    stars that are gravitationally bound according to SUBFIND at $z=0$ and within $6R_{\mathrm{e}}$. The star-forming gas density is measured by averaging all star-forming gas cells of all snapshots in the corresponding time bins. The vertical red line represents the infall time of the galaxy into its $z=0$ host, which occurred 6.1 Gyr ago. 
    }
    \label{gas_compact}
\end{figure*}

In Fig.~\ref{orbital_decmp}, we show the orbital probability density distribution and age distribution of the stellar orbits for a typical TNG50 galaxy. We directly used the stellar age of each stellar particle from the simulation. Based on the stellar orbit distribution, we decomposed the galaxy into two components: the orbits with $\lambda_z\ge0.8$ were taken as a dynamically cold disk, all the remaining orbits with $\lambda_z<0.8$ were considered as a dynamically hot non-disk component. This decomposition was performed exactly the same way as for the Fornax cluster galaxies in \citet{Ding2023}. We then also quantified each component exactly the same way as in \citet{Ding2023}.

We defined the luminosity fraction of the dynamically cold disk as 
\begin{align}\label{eq:fcold}
f_{\mathrm{cold}} =\sum_{k}^{\lambda_z\ge0.8} L_k / \sum L_k,
\end{align}
 where $L_k$ is the r-band luminosity of each stellar particle $k$ that belongs to orbit with certain $\lambda_z$. As the stellar cold-disk fraction of a galaxy varies with radius, we calculated $f_{\rm cold}(r<R_{\mathrm{e}})$ and $f_{\rm cold}(r<2R_{\mathrm{e}})$ with orbits within $R_{\rm e}$ and $2R_{\rm e}$, respectively. We measured the 2D effective radius, $R_{\mathrm{e}}$, by randomly projecting the simulated galaxies to the 2D sky plane.

We defined the luminosity-weighted average stellar age of the dynamically cold disk as
\begin{align}
t_{\mathrm{cold}} =\sum_{k}^{\lambda_z\ge0.8} t_k  L_k / \sum L_k,
\end{align}
where $t_k$ is the stellar age of each particle that belongs to a orbit with certain $\lambda_z$.
Similarly, we defined the luminosity-weighted average stellar age of the dynamically hot non-disky component as
\begin{align}
t_{\mathrm{hot}} =\sum_{k}^{\lambda_z < 0.8} t_k  L_k / \sum L_k.
\end{align}
We also similarly measured the average stellar age with orbits within $R_{\rm e}$.

\section{Environmental effects on TNG50 cluster galaxies}\label{sec3}
In the following, we quantify how the star-forming gas distribution and the SF of cluster galaxies are affected after infall according to the outcome of the TNG50 simulation: we first illustrate the phenomenology with a case study and then provide statistical results with all the galaxies in the 14 clusters of TNG50 at $z=0$. We hence quantify how the gas redistribution, combined with tidal stripping in the cluster, affects the formation and survival of the dynamically cold stellar disks in such galaxies.

\subsection{Gas removal and reduction in SF after infall}

\subsubsection{A case study}
We can illustrate the processes that take place according to TNG50 with an example galaxy: the TNG50 object at $z=0$ with Subhalo ID 96768 and galaxy stellar mass $10^{10.5}$ \msun. This is a (FoF) satellite of the massive halo with FoF ID 2 at $z=0$ (with $M_{200c} = 10^{13.8}$ \msun)
and crossed its virial radius for the first time about 6 Gyr ago (Fig.~\ref{def_infall}).

In the top panel of Fig.~\ref{gas_compact}, we see that this galaxy had a persistently high SFR before falling into the cluster. About one billion years after infall, the SFR sharply decreased, even though a long tail of SF at very low rates continues to today.

To understand the assembly history of this galaxy's structure, we further examined the spatial distribution of gas and stars across cosmic epochs.  We first traced back the main progenitor of the galaxy and took the average star-forming gas density spatial distribution at all snapshots within five different epochs, defined by the ranges [0, 2.5, 4, 6, 10, 14] Gyr.
We then compared such gas maps with the $z=0$ spatial distribution of the stellar particles in the present-day galaxy, separated into five bins according to their stellar age, and constructed the surface mass density of stars in each bin.

As shown in the middle panel of Fig.~\ref{gas_compact}, the star-forming gas was concentrated in the very inner regions at ancient times and later became more spatially extended, forming stars in a disk: based on the maps in the bottom panels, the stellar disk presents an inside-out growth. However, after the galaxy fell into the cluster, about $\sim 6$ Gyr ago, the gas in the outer disk was quickly removed and the remaining gas shrank to the inner regions of the system, leaving a central SF lasting until $z=0$: this post-infall phase exhibits an outside-in quenching. 

\begin{figure*}
    \centerline{
        \includegraphics[width=2.0\columnwidth]{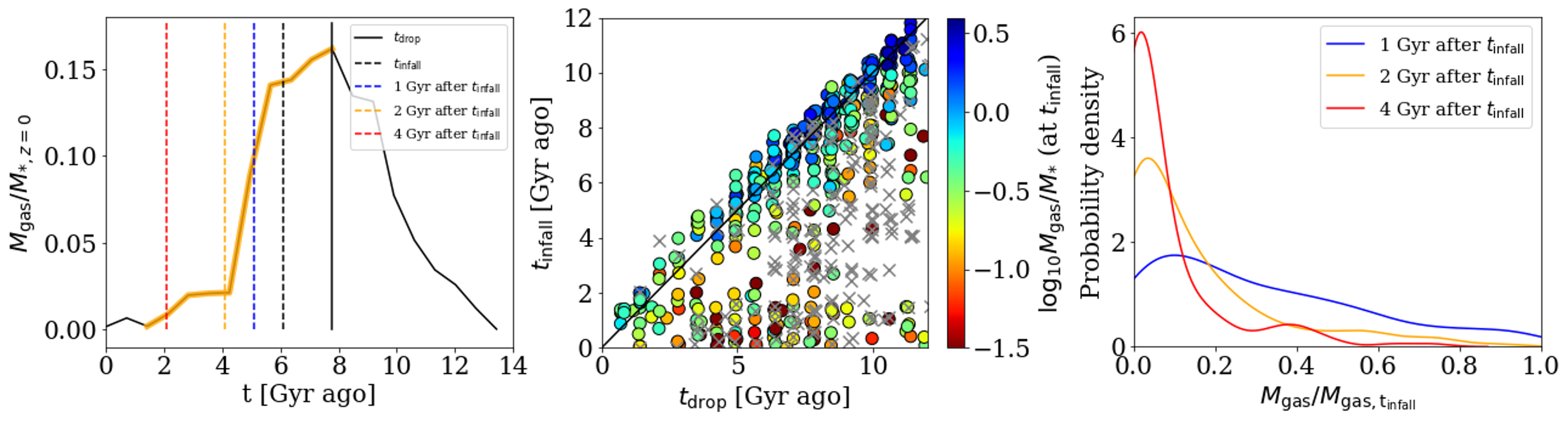}}
    \caption{{Reduction in star-forming gas mass in TNG50 cluster satellites after infall.} {Left:} Evolution of the star-forming gas mass of the example galaxy of Fig.~\ref{gas_compact} as a function of lookback time in gigayears. The yellow region shows a clear steep drop in the star-forming gas mass. The solid black line represents the starting point of such a gas drop, which we define as $t_{\mathrm{drop}}$. The dashed black line represents $t_{\rm infall}$ as defined in Section~\ref{SS:tinfall}. The dashed blue, orange, and red lines represent 1 Gyr, 2 Gyr, and 4 Gyr after $t_{\mathrm{drop}}$, respectively. {Middle:}  Relationship between the infall time and the gas drop time for all selected TNG50 cluster satellites: each point represents one galaxy, colored by the mass fraction of star-forming gas in the galaxy at infall. Gray crosses represent preprocessed galaxies that are otherwise excluded from our analysis. The solid black line represents the one-to-one line.  {Right:} Probability density distribution of the ratio of remaining star-forming gas mass compared to that at infall, for all the selected TNG50 galaxies inspected 1 Gyr (blue), 2 Gyr (orange), and 4 Gyr after $t_{\mathrm{drop}}$ (red).}
    \label{gas_drop}
\end{figure*}

In fact, as shown in the bottom panels, the oldest stars in the present-day galaxy are concentrated in the inner regions, whereas slightly younger stars (but still formed before infall) are distributed in a spatially extended disk. Stars formed after infall appear more and more centrally concentrated. The current spatial distributions of stars with different ages are consistent with the star-forming gas distributions across time. 

\subsubsection{Drop in the gas mass}
We quantified the time evolution of the star-forming (i.e., cold; see Sect.~\ref{SS:gasprops}) gas mass in all selected TNG50 cluster galaxies. 

For each galaxy, we find its main progenitor across all the past snapshots and extract the mass of star-forming gas therein. As an example, in the left panel of Fig.~\ref{gas_drop}, we show the evolution of the star-forming gas mass in the main progenitor of the galaxy of Fig.~\ref{gas_compact} (Subhalo ID 96768) as a function of cosmic time. The star-forming gas in this galaxy increased steadily from ancient epochs and until about 8 Gyr ago, then it started to decrease and only a few percent of its peak mass is left at $z=0$. 

We defined the time of ``gas mass drop'' ($t_{\mathrm{drop}}$)
by searching for a pair of adjacent local maximum and minimum along the star-forming gas mass evolution curve. We chose the largest jump and call the local maximum $t_{\mathrm{drop}}$. 
The local maximum we found is just the global maximum of the curve in 93\% of the cases. In 7\% of the cases, there could be another drop in gas caused by a merger or other physical processes that occurred long before it fell into the cluster; we excluded this drop and chose the one that occurred close to its infall into the cluster.

We extracted $t_{\mathrm{drop}}$ for all the selected TNG50 galaxies and find it to be typically strongly correlated with the time of infall, $t_{\mathrm{infall}}$, as shown in the middle panel of Fig.~\ref{gas_drop}. The gas drop happens close to the falling into a cluster, especially for galaxies that are gas-rich at infall (see the color bar).
For galaxies that have already consumed or lost most of their gas before infall, the largest gas drop we identify is not associated with the infall, but to physical processes happened earlier on. Similarly, for the preprocessed galaxies (gray crosses), the largest gas drop also occurred before infall into the current cluster. 

To quantitatively examine how fast the star-forming gas decreases in the cluster environment, we measured the ratio of satellite's star-forming gas mass at 1 Gyr, 2 Gyr, and 4 Gyr after infall in comparison to the star-forming gas mass at infall: $M_{\rm gas}/M_{\mathrm{gas,}t_{\rm infall}}$. The distributions of these retained star-forming gas ratios are shown in the right panel of Fig.~\ref{gas_drop}. 
About 1 Gyr after infall, about 42\% of galaxies have only $\lesssim 20\%$ of their star-forming gas remaining, 76\% of galaxies have $\lesssim 50\%$ of their star-forming gas remaining, and 93\% of galaxies have $\lesssim 80\%$ of their star-forming gas remaining. About 2 Gyr after infall, the number of galaxies having only $\lesssim 20\%$, $\lesssim 50\%$, or $\lesssim 80\%$ of the remaining star-forming gas increase to 73\%, 92\%, and 99\%. After 4 Gyr after infall, almost 90\% of the galaxies are left with less than $20\%$ of the gas mass they had at infall.

\begin{figure*}
    \centerline{
        \includegraphics[width=2.0\columnwidth]{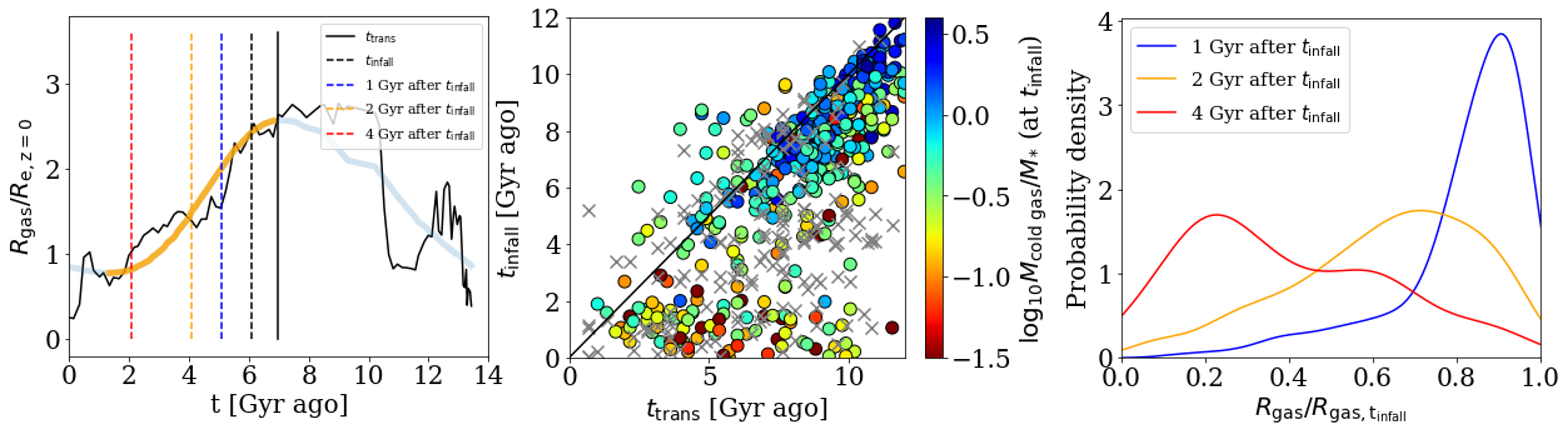}}
    \caption{{Reduction in the spatial extent of the star-forming gas in TNG50 cluster satellites after infall.} {Left:} Evolution of the SF-weighted gas radius of the example galaxy of the previous figures, as a function of lookback time in gigayears. The black curve represents the outcome along the main progenitor branch of the galaxy; the blue curve represents a Fourier-smoothed version thereof. The yellow curve highlights the time when the gas radius gets smaller. The solid vertical black line represents the time when the gas radius starts to shrink, which we define as $t_{\mathrm{trans}}$. The dashed black line represents infall, $t_{\rm infall}$. The dashed blue, orange, and red lines represent 1 Gyr, 2 Gyr, and 4 Gyr after $t_{\mathrm{trans}}$, respectively. {Middle:} Relationship between infall time and the time when the gas radius starts to shrink for all selected TNG50 cluster satellites, colored by the mass of star-forming gas at infall. Gray crosses represent preprocessed galaxies. The solid black line represents the one-to-one line.  {Right:} Probability density distribution of the gas radius ratio to that at infall for all selected TNG50 galaxies inspected 1, 2, and 4 Gyr after $t_{\rm infall}$, respectively.
    }
    \label{gas_shrinking}
\end{figure*}
\subsubsection{Shrinking of the star formation to the inner regions}
\label{SS:gas_shrinking}
When interacting with the cluster environment, not only the amount of gas, but also its spatial distribution changes, as we show next. Thus, not only the SFR, but also the spatial distribution of the star-forming regions change: all this ultimately also affects the formation of the dynamically cold stellar disks. 

We quantified the spatial distribution of the star-forming gas in a galaxy via the gas radius, $R_{\rm gas}$, defined as the SFR-weighted mean radius of the star-forming gas.
For each galaxy, we measured this radius along its main progenitor branch. In the left panel of Fig.~\ref{gas_shrinking}, we show the evolution of the gas radius of our example galaxy (Subhalo ID 96768 in TNG50 at $z=0$) as a function of cosmic time. The gas radius in this galaxy increased from ancient times to about 7 Gyr ago, then it started to decrease and shrank to approximately the stellar $R_{\rm e}$ (see Sect.~\ref{SS:decomposition}) of the galaxy at $z\sim 0$.

Similar to the definition of $t_{\mathrm{drop}}$, we defined a transition time for the gas radius $t_{\mathrm{trans}}$
by searching again for adjacent pairs of local maximum and minimum along a smooth curve of the gas-radius evolution. We applied a Fourier smoothing to avoid local noise. We chose the largest decrease from the local maximum to local minimum along the evolution curve as the time when the galaxy's gaseous structure starts to shrink. For the case shown in the figure, we find that the gaseous radius started to shrink about $t_{\mathrm{trans}}=7$ Gyr ago.

By measuring the $t_{\mathrm{trans}}$ for all selected TNG50 cluster satellites, we find again a correlation between $t_{\mathrm{trans}}$ and $t_{\mathrm{infall}}$, as shown in the middle panel of Fig.~\ref{gas_shrinking}. This correlation is in place for the galaxies that carried a significant amount of star-forming gas at infall, whereas it is weak or vanishing for galaxies that have little star-forming gas at infall and for the preprocessed galaxies.

To quantitatively examine the rate at which the gas radius of TNG50 cluster satellites shrinks, we measured the ratio between the gas radius at any given time and the gas radius at infall: $R_{\rm gas}/R_{\mathrm{gas,}t_{\rm infall}}$. We show the distributions of such a ratio for all selected galaxies 1 Gyr, 2 Gyr, and 4 Gyr after infall in the right panel of Fig.~\ref{gas_shrinking}.
For all TNG50 cluster satellites, the gas radius becomes smaller after infall. 

At 1 Gyr after infall, about 1\%, 7\%, and 31\% of 
galaxies have their star-forming gas extents reduced to 20\%, 50\%, and 80\% of the original size.
About 2 Gyr after infall, the number of galaxies
having $R_{\rm gas}/R_{\mathrm{gas,}t_{\rm infall}} \lesssim 0.2$, $R_{\rm gas}/R_{\mathrm{gas,}t_{\rm infall}} \lesssim 0.5$, and $R_{\rm gas}/R_{\mathrm{gas,}t_{\rm infall}} \lesssim 0.8$ increase to 5\%, 27\%, and 74\%.
At 4 Gyr after infall, the corresponding fractions of galaxies increase to 25\%, 66\%, and 92\%, respectively.

The reduction in both the mass and the spatial extent of the star-forming gas of a satellite is consistent with this gas being stripped by RPS. In fact, as mentioned in the Introduction, thousands of jellyfish galaxies could be identified in TNG50 groups and clusters across cosmic epochs \citep{zinger2023, rohr2023}.

\begin{figure}
    \centerline{
        \includegraphics[width=1.\columnwidth]{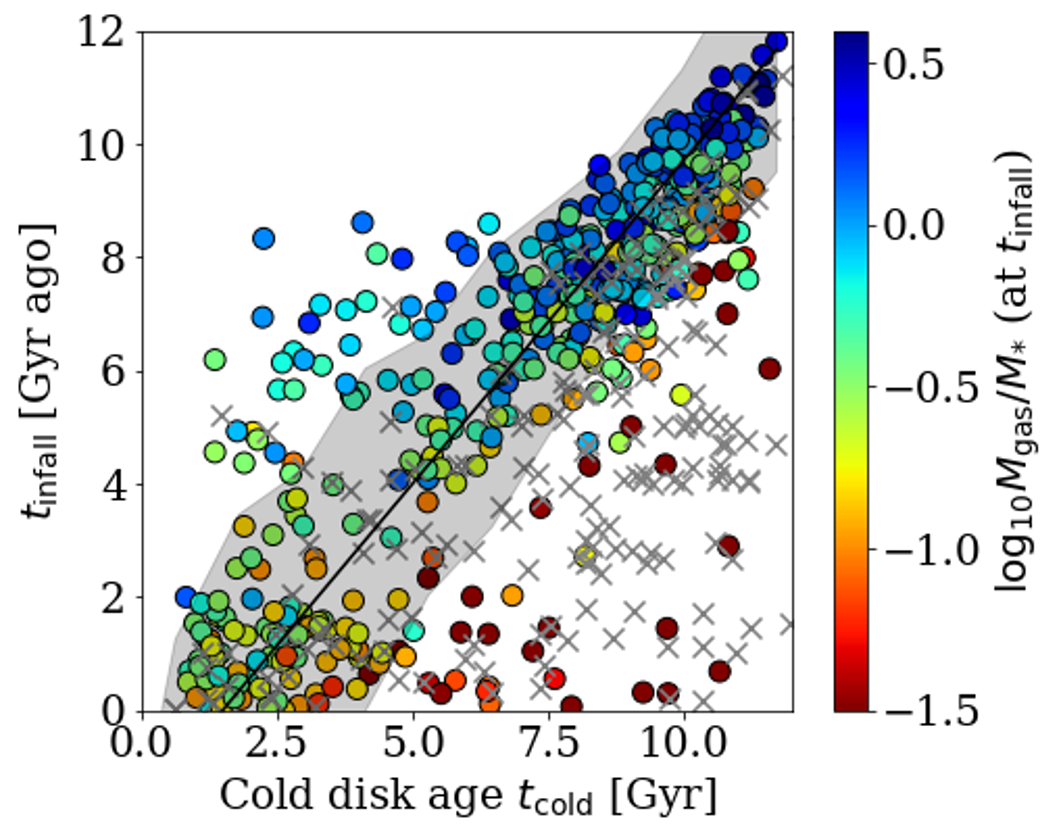}}
    \caption{{Infall time versus average stellar age of the dynamically cold disk in all selected TNG50 cluster galaxies}. Each point represents one satellite, colored by the star-forming gas content at infall. Gray crosses represent preprocessed galaxies. The black line gives the linear fit $t_{\mathrm{infall}} = 1.14t_{\mathrm{cold}} -1.71$ Gyr with $1\sigma$ scatter of about 2.
    }
    \label{infall_vs_disk_age_all_gas_content}
\end{figure}

\begin{figure*}
    \centerline{
        \includegraphics[width=1.8\columnwidth]{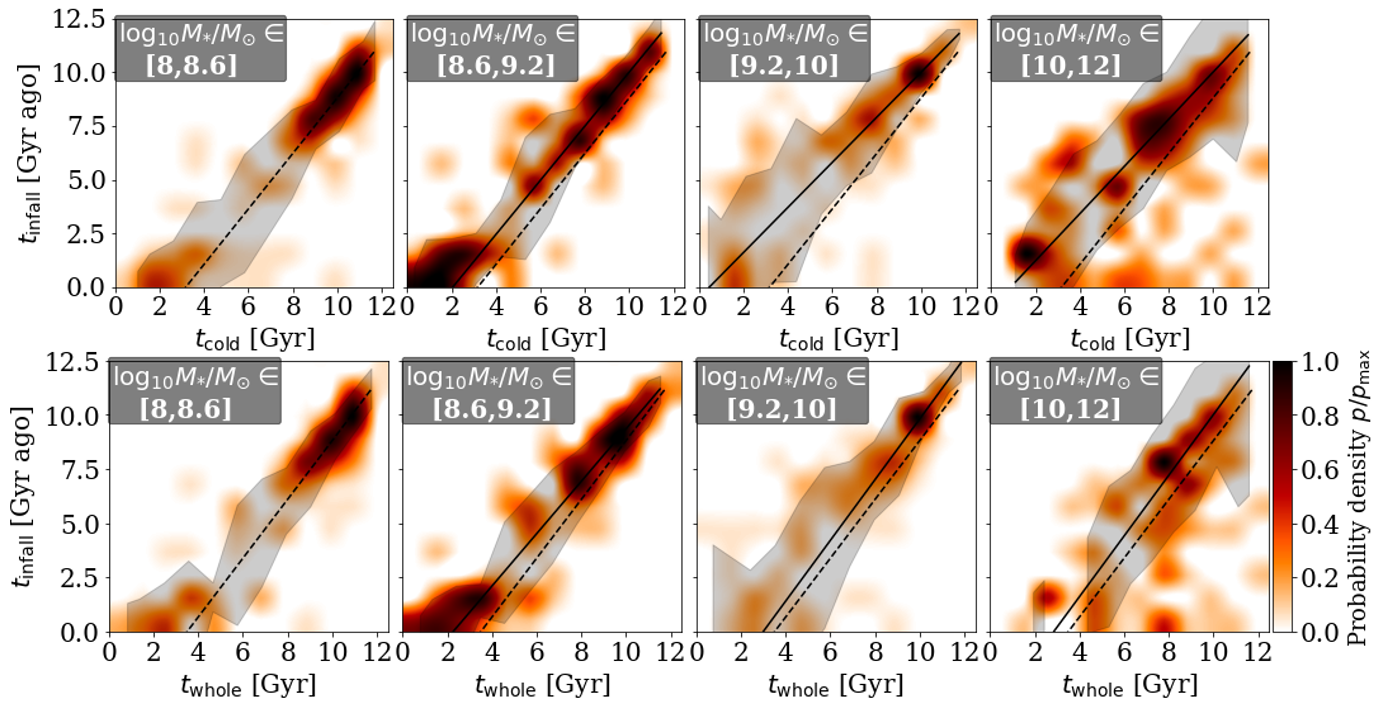}}
    \caption{{Infall time versus average stellar age of the dynamically cold disk in TNG50 cluster galaxies as a function of galaxy stellar mass.} We show the density contours of infall time versus stellar age of the dynamically cold disk {\em (top panel)\/} and versus the average stellar age of the whole galaxy age {\em (bottom panel)\/}. Different galaxy mass bins, $\log_{10}M_{*}/\mathrm{M_\odot}\in$[8,8.6], [8.6,9.2], [9.2.10], and [10,12] are shown from left to right. The dashed line in the first mass bin [8,8.6] represents its linear fit, repeated for comparison for the higher-mass galaxies. Shaded regions represent the galaxy-to-galaxy variation, i.e., the scatter to the linear fit.
    }
    \label{infall_vs_disk_age_mass_bin}
\end{figure*}

\subsection{Stellar age of the dynamically cold disk as a proxy of infall}
\label{SS:tcold}
As seen in the previous sections, the falling of a galaxy into a cluster implies significant changes to (i.e., reductions in) the amount and spatial distribution of its star-forming gas. In turn, such decreases are expected to affect the SF history of a galaxy and hence the properties of a possible dynamically cold stellar disk. For a real galaxy at $z=0$, we cannot directly observe the evolution history of the star-forming gas, but we can derive stellar population properties of the different stellar structural components of the galaxy via IFU observations. 

In Fig.~\ref{infall_vs_disk_age_all_gas_content}, we show the relationship between infall time and the average stellar age of the dynamically cold disk at $z=0$, for all selected TNG50 cluster galaxies. The two quantities are highly correlated, with ancient infallers having on average older stellar disks and recent infallers having younger disks.
We fit a linear line of $t_{\mathrm{infall}} = 1.14t_{\mathrm{cold}} -1.71$ Gyr to the TNG50 results; with a $1\sigma$ scatter of about 2 Gyr. Therefore, the stellar age of the dynamically cold disk in cluster galaxies can serve as a reasonable proxy for the time of infall.

The gray crosses represent preprocessed galaxies (excluded from the fit above) that fell into another cluster or group before joining their $z=0$ host. They have typically already stopped forming stars at this point.

The outliers below the correlation had little star-forming gas at infall, whereas those above the correlation had a lot of star-forming gas at infall: for those, SF in a dynamically cold disk progressed for $\gtrsim 2$ Gyr after infall. The youngest (i.e., bluest) outliers with significantly young cold disks are mostly galaxies in lower-mass clusters. In fact, although we have chosen hosts within a narrow halo mass range, and even though the correlations between $t_{\rm infall}$ and $t_{\rm cold}$ are similar in all the 14 clusters taken individually, the majority of the outliers are gas-rich galaxies falling in hosts at the lower end of our range (groups, in fact) or in hosts that are not completely relaxed.

To check the dependence of this correlation on galaxy mass, we further divided all cluster satellites into four mass bins: $\log_{10}M_{*}/\mathrm{M_\odot}\in(8,8.6),\ (8.6, 9.2), \ (9.2,10)$, and \ $(10,11)$ (see Fig.~\ref{infall_vs_disk_age_mass_bin}). In the top panels, we show $t_{\mathrm{infall}}$ versus $t_{\mathrm{cold}}$ as in Fig. Fig.~\ref{infall_vs_disk_age_all_gas_content}. We fit a linear line to the correlation in each mass bin and find that more massive galaxies lie to the left of the correlation for the less massive galaxies -- SF in dynamically cold disks can last for about 1-2 Gyr longer after infall in the more massive galaxies ($M_* \gtrsim 10^9$ \,$\Msun$) than in the less massive ones. The $1\sigma$ scatter reads 2.3, 1.9, 2.1, 2.9 billion years, from the low to high mass bins, respectively. The galaxy-to-galaxy variation is largest at the highest-mass end -- there are old outliers (red markers) with only little star-forming gas at infall, thus lying below the correlation. It should be remembered that, in this mass range, active galactic nucleus feedback is also of relevance and can both deplete the gas and stop SF in addition or simultaneously to environmental effects \citep{donnari2021a}.

We also examine the correlation between infall time and the stellar age of the whole galaxy, $t_{\mathrm{whole}}$, in the bottom panel of Fig.~\ref{infall_vs_disk_age_mass_bin}: this is the light-weighted average stellar age of the whole galaxy. The relations are similar to those discussed above, but with weaker trends as a function of galaxy mass.

\subsection{Lower fractions of the dynamically cold disk in ancient infallers}
\label{SS:fcold}
The gas removal in the cluster environment affects the SF history (and hence the stellar populations) of galaxies, especially in the stellar dynamically cold disk. Consequently, we expect that, in turn, also the luminosity or mass fraction of the stellar cold disk in galaxies changes because of the environment, and hence after infall. This has been shown with theoretically motivated disk measures in IllustrisTNG by \citealt[][]{Joshi2020, GalandeAnta2022} (see Sect.~\ref{sec:intro}), who showed that tidal processes are also at play. The quantity we can derive from observations is the cold-disk luminosity fraction in galaxies at $z=0$, which is a combined result of the above-mentioned processes. We then derived the cold-disk fraction (as per Sect.~\ref{SS:decomposition}) of each selected TNG50 cluster galaxy at infall. 

For this, we adopted the simulation output from the snapshot along the main progenitor branch of each galaxy that is closest to, and precedes, the infall time into the cluster. At this time, for non preprocessed galaxies and typically, the stellar structures of future cluster members have not been significantly affected yet by the interaction with the cluster, thus representing their property as centrals. We took the snapshot at $z=0$ to derive each galaxy's present-day cold-disk fraction.
To make the comparison between $z=0$ and at infall, we took the mass-weighted (instead of the luminosity-weighted) cold-disk fraction: for each galaxy, this was calculated as the mass fraction of the cold orbits ($\lambda_z\ge0.8$) within $1R_{\mathrm{e}}$ and $2R_{\mathrm{e}}$, respectively, similar to the definition of the luminosity fraction in Eq.~\ref{eq:fcold} but with the luminosity replaced by mass. As the galaxy size may change with time, we used the effective radius $R_{\mathrm{e}}$ at $z=0$ for the calculation of the cold-disk fractions, both at infall and at $z=0$, to keep investigating the ``same'' regions.

\begin{figure*}
\begin{center}
    \includegraphics[width=1.8\columnwidth]{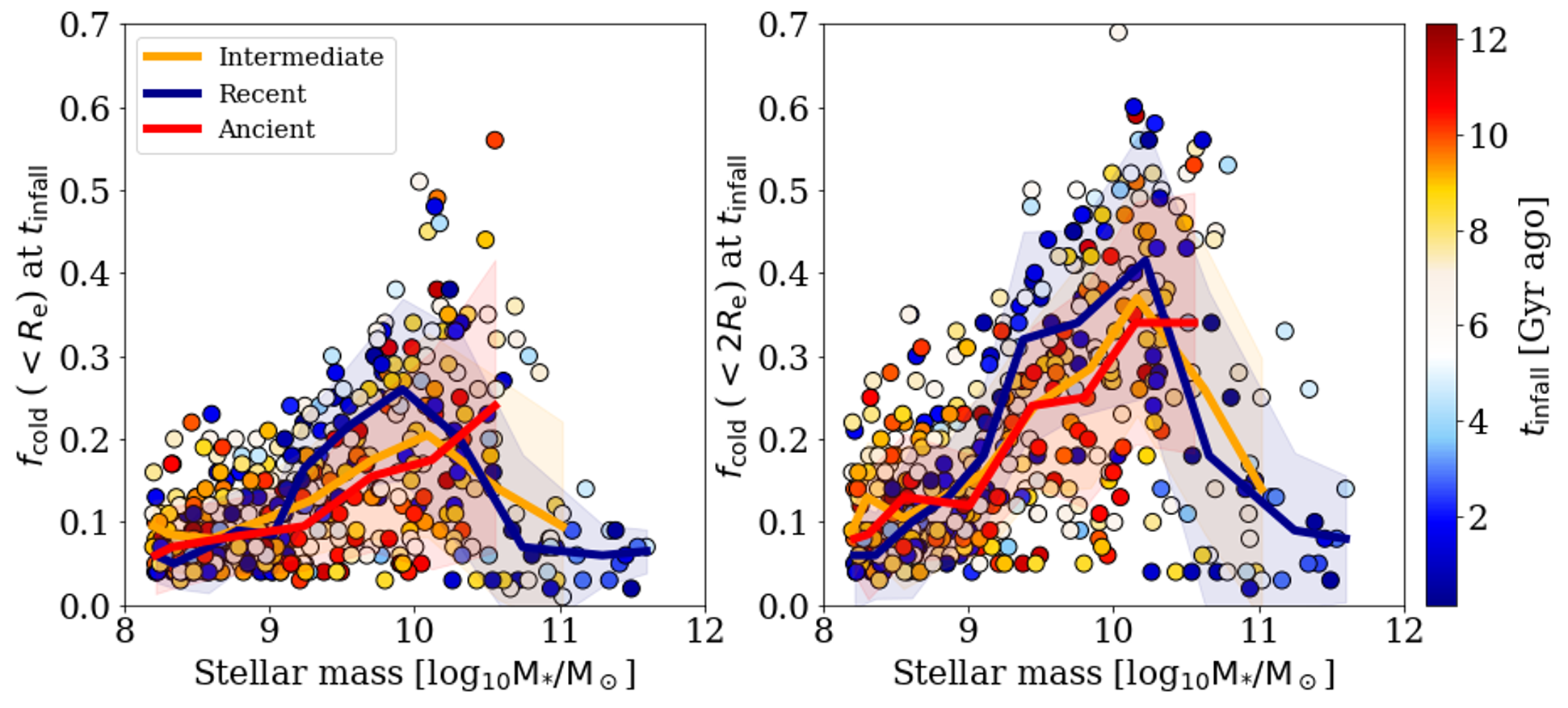}
    \includegraphics[width=1.8\columnwidth] {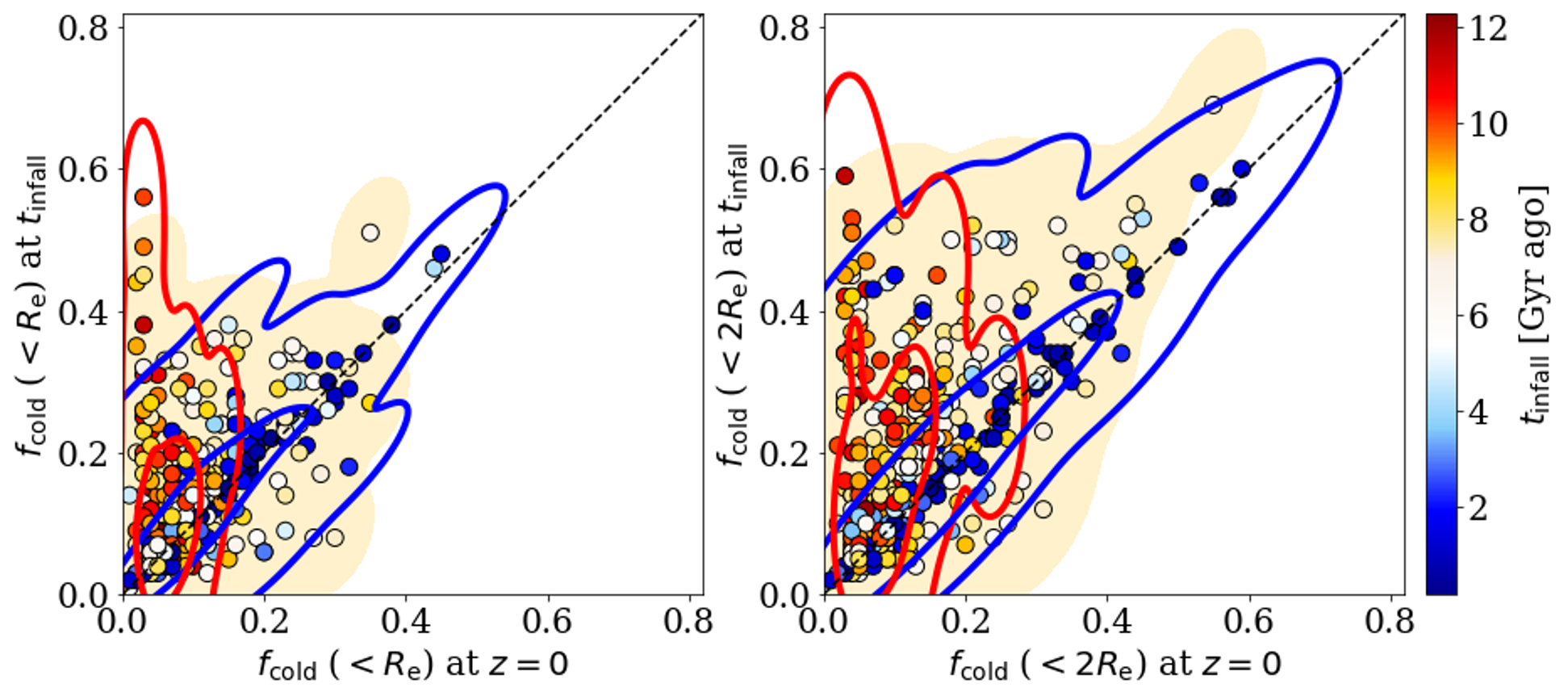}
    \includegraphics[width=1.8\columnwidth]{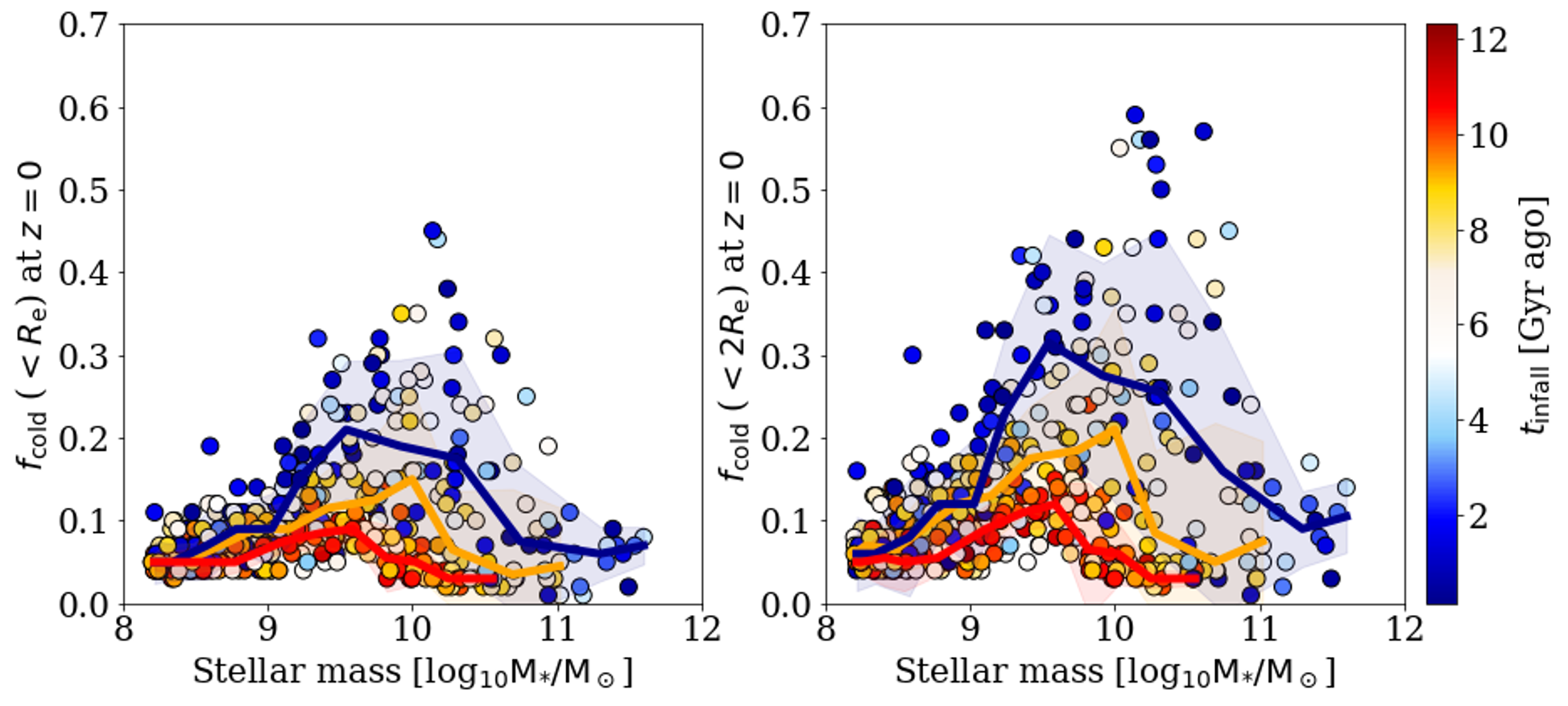}
    \caption{
        {Stellar mass fraction in the dynamically cold disks in cluster galaxies according to TNG50, at infall and at the present day.} In all rows, we show the mass-weighted cold-disk fraction of galaxies within $1R_{\rm e}$ (left) and within $2R_{\rm e}$ (right), whereby the effective radius is obtained at $z=0$. In all panels, we color code the markers by the time since infall, according to the color bars. In the top and bottom rows, red, yellow, and blue curves and shaded regions denote population medians and 1-$\sigma$ variations of the stellar cold-disk fraction in bins of present-day galaxy stellar mass for ancient, intermediate, and recent infallers, respectively. In the middle row, the fraction in the dynamically cold disks is compared between the infall time and the present day, with the red and blue contours representing the 1$\sigma$ and 3$\sigma$ density contours of the ancient and recent infallers, respectively, and with the yellow shaded region denoting the 3$\sigma$ density contour of intermediate infallers. As clearly shown by the middle plots, the stellar mass fraction of cold, disky orbits at $z=0$ is smaller for galaxies that fell into their $z=0$ cluster longer ago, i.e., the time spent within the cluster environment correlates with a reduction in stellar disk structures.}
    \label{fcold_vs_mass_all}
\end{center}
\end{figure*}

In Fig.~\ref{fcold_vs_mass_all}, we show the distributions of the cold-disk fractions in cluster satellites according to TNG50 at the time of infall, $t_{\rm infall}$ (top row), and at the present time (bottom row), and compare the two in the middle row. In all panels, we keep track of the infall time into their $z=0$ cluster and highlight trends for ancient, intermediate, and recent infallers as per Sect.~\ref{SS:tinfall}. The two columns refer to two different galaxy apertures within which the cold-disk fraction is measured.

In general, the cold-disk fraction at infall (i.e., prior to the bulk of environmental processes) is maximum in galaxies with stellar mass $M_{\ast} \sim 10^{10}$\msun and decreases toward both the low-mass and high-mass ends. The cold-disk fraction peaks at $M_{\ast} \sim 10^{9.5}-10^{10.5} M_\odot$ with $f_{\mathrm{cold}}(r<R_\mathrm{e}) =0.24$, 0.20, and 0.19 and $f_{\mathrm{cold}}(r<2R_\mathrm{e}) =0.36$, 0.31, and 0.29, for recent, intermediate, and ancient infallers, respectively. Namely, the cold-disk fractions in galaxies at infall are statistically lower for ancient infallers than in recent infallers, which we believe is a result of the removal or consumption of gas at earlier times.

\begin{figure*}
   \centerline{
       \includegraphics[width=1.5\columnwidth]{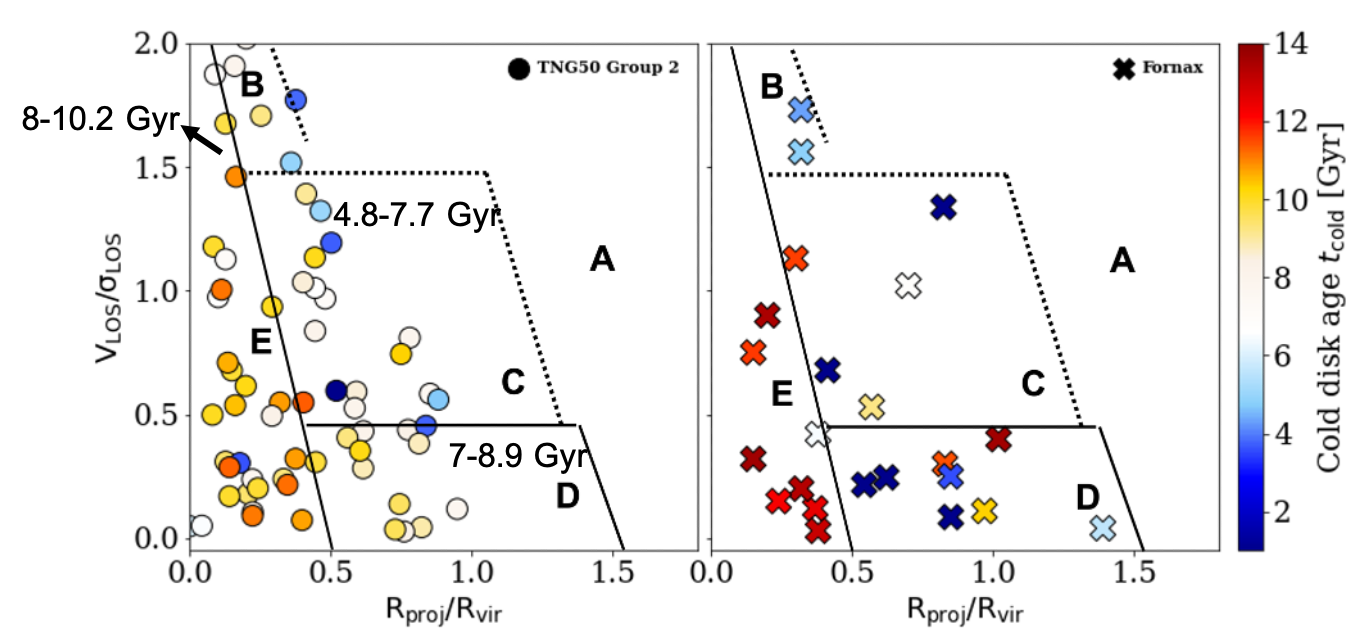}}
       \caption{{Distribution of TNG50 and observed cluster galaxies in the phase-space of projected LOS velocity versus projected cluster-centric radius, color coded by the stellar age of the dynamically cold component}. In particular, we compare Fornax cluster members (right) to cluster galaxies of the TNG50 halo with Group ID 2 (left), in a random projection. The boundaries of regions A, B, C, D, and E are defined as in \citet{Rhee2017}. Both simulated and observed cluster galaxies exhibit similar trends, with galaxies closer to the cluster center typically having older stars in their disk components.
   The annotated numbers give the 25th-75th percentile ranges of the true infall time of the simulated galaxies in region E, D, and B$+$C, respectively.
   }
   \label{cluster_phase_space_v2}
\end{figure*}

In fact, as introduced earlier, the stellar cold disk of satellites in groups and clusters may be disrupted by tidal forces, and more precisely by tidal shocking, after the galaxies fall into their host \citep{Joshi2020,GalandeAnta2022}. We will not repeat the analysis of how the tidal disruption occurs, but we can quantify the effect by comparing the cold-disk fraction at $z=0$ and that at $t_{\rm infall}$, measured within $1R_{\mathrm{e}}$ and $2R_{\mathrm{e}}$, respectively, in the middle panels of Fig.~\ref{fcold_vs_mass_all}.
On average, the fraction of the cold disk within $R_{\rm e}$ ($2R_{\rm e}$) at $z=0$ is 0.90, 0.66, and 0.47 (0.86, 0.60, and 0.39) times of those at infall for recent, intermediate, and ancient infallers, respectively. Namely, across the satellite population, the cold-disk fractions at $z=0$ are lower than at infall, for most galaxies. For recent infallers, the cold disk fractions within $2R_{\mathrm{e}}$ have not been significantly altered after falling into the cluster; for intermediate infallers, their cold-disk fractions within $R_{\rm e}$ are weakly affected, whereas disk fractions within $2R_{\rm e}$ are moderately affected; on the other hand, environment
significantly affects the fraction of the cold disk at all radii for ancient infallers. 

By comparing the top and bottom rows of Fig.~\ref{fcold_vs_mass_all}, we see that, for recent and intermediate infallers, the cold-disk fractions at $z=0$ exhibit a similar trend as a function of stellar mass as at infall, but are systematically lower, peaking at about $f_{\mathrm{cold}}(r<R_\mathrm{e}) = 0.19$ and $0.13$ ($f_{\mathrm{cold}}(r<2R_\mathrm{e}) =0.29$ and $0.17$) for recent and intermediate infallers, respectively, in galaxies with $M_* = 10^{9.5-10.5}$\,$\Msun$.
The difference between recent and intermediate infallers at $z=0$, especially for the cold-disk fractions in the inner $R_\mathrm{e}$, seems to partly inherit the already-in place difference at the time of infall. For ancient infallers, on the other hand, the cold-disk fraction becomes very low at $z=0$, with an average of $f_{\mathrm{cold}}(r<R_\mathrm{e}) = 0.06$ and $f_{\mathrm{cold}}(r<2R_\mathrm{e}) =0.08$ for selected cluster galaxies across the mass range. 

Our results are generally consistent with previous results that the mass fraction of disks decreases after a few gigayears fall into a cluster. However, it seems that the inner and outer regions of the disks are affected by tidal forces with different timescales. 
We find that the disks within $R_\mathrm{e}$ are only weakly affected in recent infallers ($t_{\rm infall} <4$ Gyr ago), moderately affected in intermediate infallers ($4<t_{\rm infall} <8$ Gyr ago), and significantly disrupted in ancient infallers ($t_{\rm infall} >8$ Gyr ago). The outer disks are interrupted a bit earlier, as shown in the middle panels of Fig.~\ref{fcold_vs_mass_all}, when considering the whole disks at $r<2R_\mathrm{e}$, they are already significantly affected in quite a few recent infallers, more than that for the disk at $r<R_\mathrm{e}$. When considering the disk within the entire galaxy region, \citet{Joshi2020} found that the disk is significantly disrupted in the cluster, with the timescale of transition from disks to non-disks being 0.5-4 Gyr.

\section{The Fornax cluster versus a TNG50 analog}\label{sec4}

Given the analysis described in the previous sections, we can now see whether the general picture predicted by the TNG50 simulation corresponds to reality. To this aim, we compared galaxies in the Fornax cluster to those in an analog cluster selected among the TNG50 groups and clusters at $z=0$. In particular, we used the TNG50 halo with Group ID 2 as the comparison system: with a total virial mass of $M_{200c}=10^{13.81}$\msun and a virial radius of $R_{200c}=0.85$ Mpc, it is closest in mass and size to the real Fornax cluster ($M_{200c}=10^{13.85}$ \msun and $R_{200c}=0.7$ Mpc). Throughout this paper, all the parameters and properties of Fornax galaxies are taken directly from \citet{Ding2023}.

\subsection{The stellar age of the dynamically cold disk}

\subsubsection{Cold disk age versus infall time}
In Sect.~\ref{SS:tcold} we show that, according to TNG50, there is a tight correlation between infall time and the stellar age of the dynamically cold disk. We checked if it is true also in the Fornax cluster. 

From observations, it is not straightforward to obtain the infall time of galaxies; instead, galaxies' location in the phase space of projected line-of-sight (LOS) velocity versus the cluster-centric radius is statistically used as an indicator \citep{Rhee2017}. To make it directly comparable with the Fornax cluster, we randomly projected the TNG50 Group 2 into the 2D observational plane.

In Fig.~\ref{cluster_phase_space_v2}, we show all the galaxies in the phase space of the projected LOS velocity versus the cluster-centric radius, colored by the stellar age of their cold disks: the projected LOS velocity is normalized by the velocity dispersion of the cluster $V_{\rm LOS}/\sigma_{\rm LOS}$, and the projected cluster-centric radius is normalized by the virial radius of the cluster $R_{\rm proj}/R_{\rm vir}$.  
We divided the phase space into multiple regions following \citet{Rhee2017}, so that, statistically, galaxies in region E have a higher probability of being ancient infallers, whereas galaxies in regions B, C, and D have a higher probability of being recent infallers. This is indeed the case also for the TNG50 cluster with Group ID 2: the typical infall time of satellites in region E is 8-10.2 Gyr ago (25th-75th percentiles), whereas that of satellites in region; for example, B$+$C is 4.8-7.7 Gyr ago.

We find that, in the Fornax cluster (right panel), galaxies in region E (ancient infallers) have the oldest cold disks, whereas galaxies in region B, C, and D (recent infallers) are more likely to have younger cold disks. This is generally consistent with the trends of the members of the Fornax analog selected from the TNG50 simulation (left panel), and consistent with the correlation between infall time and cold disk age uncovered in Fig.~\ref{infall_vs_disk_age_all_gas_content}. These findings support the usage of the simulation-based relationship between infall time and cold disk age to infer the former from the latter for observed galaxies, as in fact already done in \citealt{Ding2023} and as used in the following sections.

\subsubsection{Stellar age in cold versus hot stellar components}
\begin{figure}
    \centerline{
        \includegraphics[width=0.8\columnwidth]{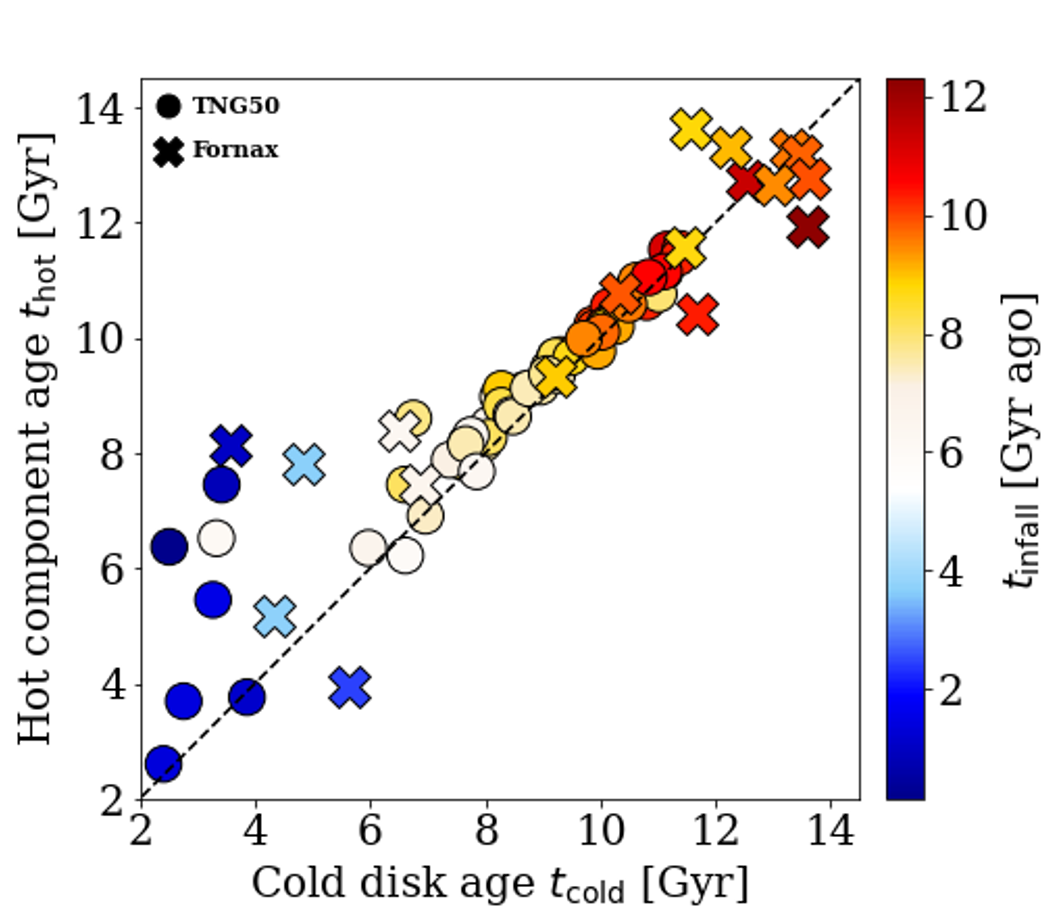}}
        \caption{{Comparison of the mean stellar age of the dynamically cold disks and of the dynamically hot non-disk components of cluster galaxies}. Circles and crosses represent galaxies from TNG50 Group 2 and from the Fornax cluster, respectively, color coded by time of infall. The dashed  black line represents the one-to-one line. Stars are similarly old in both the cold disk and hot component for ancient infallers, stars become younger in recent infallers, and are about 2 Gyr younger in the cold disk than in the hot component for the most recent infallers.
    }
    \label{cold_age_vs_hot_age_g2}
\end{figure}

To have a comprehensive view of how the infall into a cluster affects cold disk formation, we compare the stellar age of the dynamically cold disk and of dynamically hot non-disk components in Fig.~\ref{cold_age_vs_hot_age_g2}.  

In both Fornax and its TNG50 analog, stars in the cold disks are as old as those of the hot components for the ancient infallers whereas they are typically younger (up to a few billion years) than those of the hot components in recent infallers. We ascribe the similarity of the stellar ages of cold and hot components in the ancient infallers to the early gas removal at the infall.

In the Fornax cluster, the average stellar age difference between the cold disk and hot non-disk component reads: $t_{\rm hot}-t_{\rm cold}$
= $0 \pm 0.64$ Gyr, $1.3 \pm 1.4$ Gyr, and $1.7\pm1.7$ Gyr for ancient, intermediate, and recent infallers, respectively, while in the TNG50 cluster with Group ID 2, we find $0.2 \pm 0.2$ Gyr, $0.4 \pm 0.7$ Gyr, and $1.8 \pm 1.5$ Gyr, respectively. These numbers quantitatively agree with each other.

Importantly, the stellar age of the dynamically cold disks of galaxies in the field is typically $\sim 2$ Gyr younger than that of the dynamically hot bulges \citep{jin2023}, similar to the case of recent infallers in both Fornax and TNG50 clusters.

\subsection{The luminosity fractions of the dynamically cold disks}
Another major result that we have quantified on the effects of environment on galaxy structures is the correlation between the stellar (mass) fraction in cold orbits and time since infall, according to TNG50 (see Sect.~\ref{SS:fcold}). Here, we checked again if such TNG50-based predictions are consistent with the observations of Fornax. 

\begin{figure*}
\begin{center}
    \includegraphics[width=1.8\columnwidth]{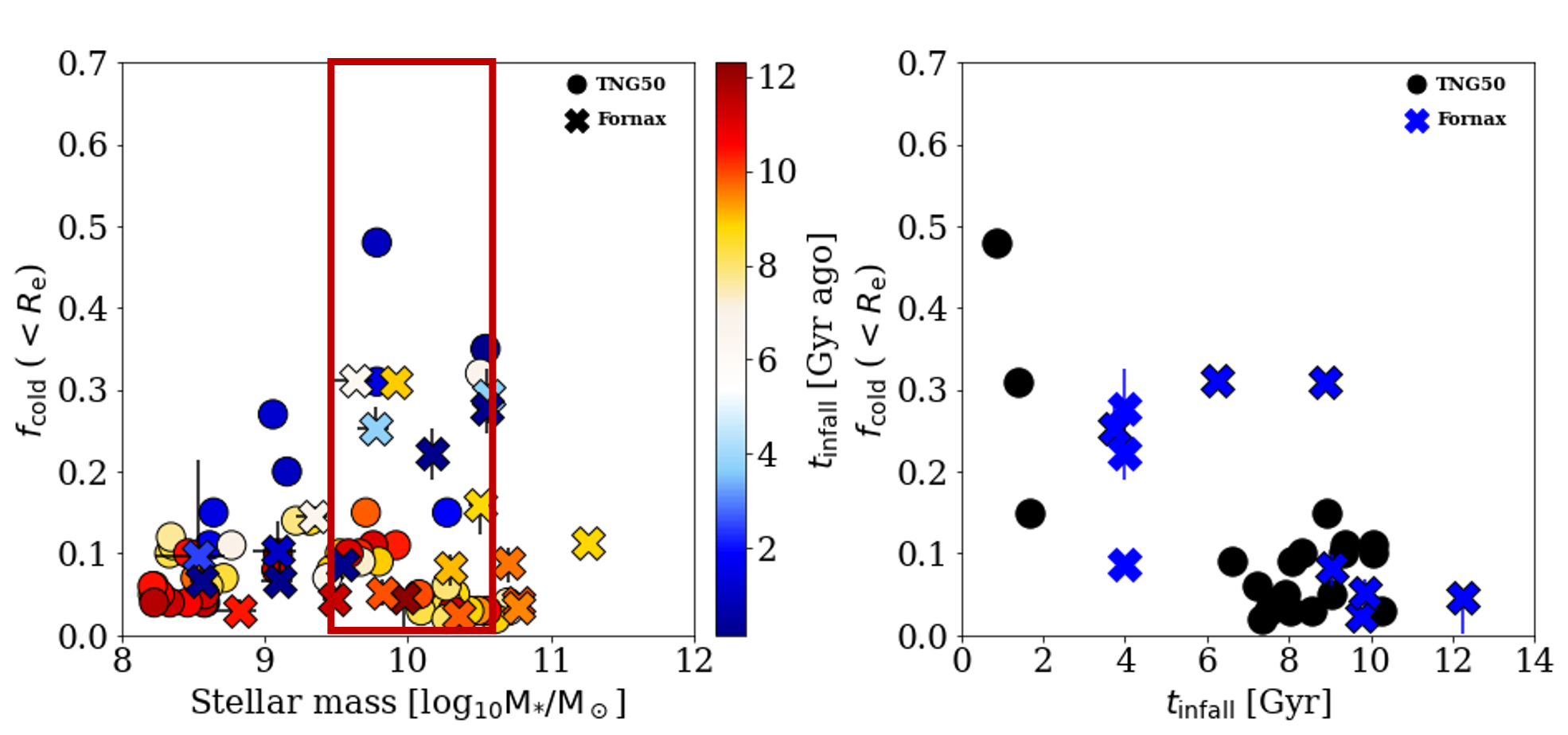}
    \includegraphics[width=1.8\columnwidth] {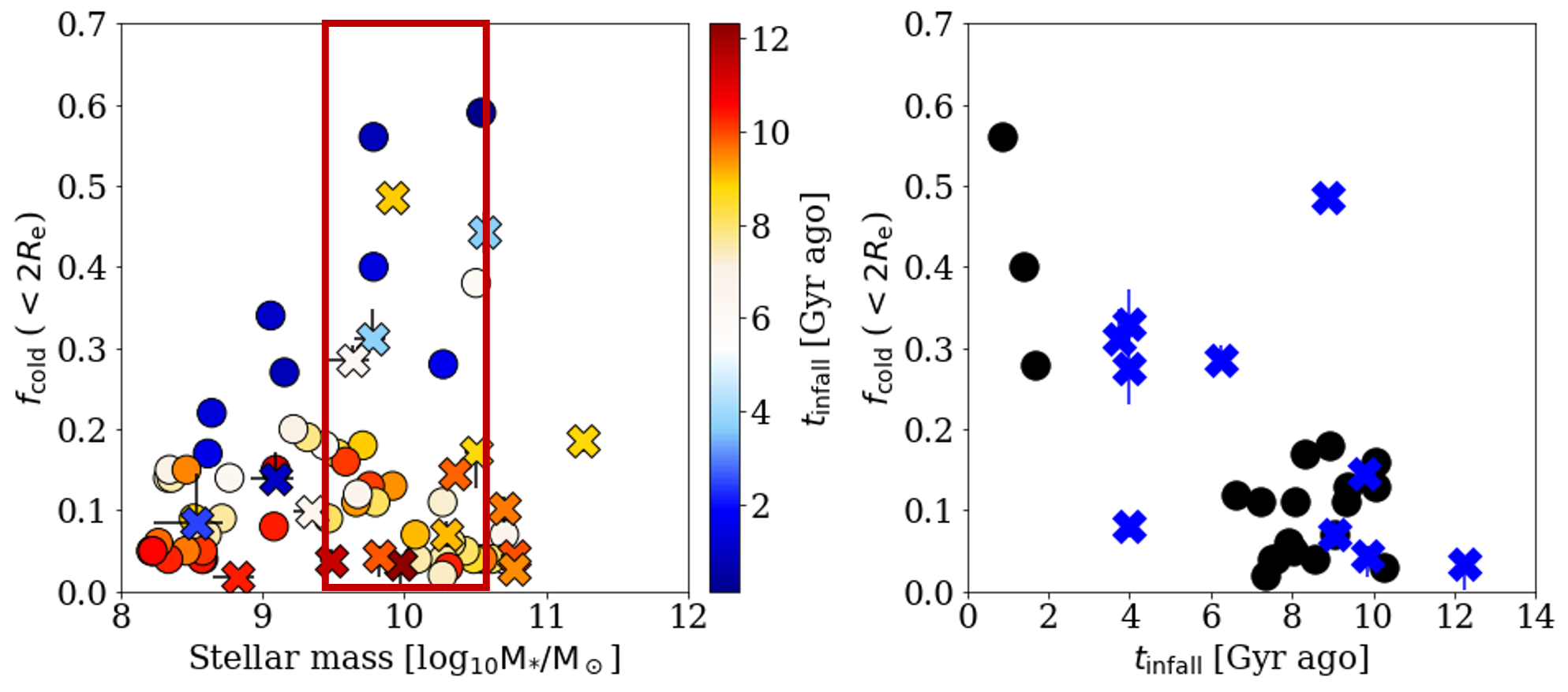}
    \caption{
        {Relationship between present-day light-weighted cold-disk fraction, galaxy stellar mass and infall time of Fornax galaxies (crosses) and cluster satellites of a TNG50 Fornax analog (circles).} The left panels are similar to those in the bottom row of Fig.~\ref{fcold_vs_mass_all} but for only two clusters (Fornax and its TNG50 analog), whereby the fraction of disky orbit is estimated within $1R_{\rm e}$ (upper left) and within $2R_{\rm e}$ (bottom left). The right panels show the present-day light-weighted cold-disk as a function of galaxies' infall time, but only in the mass range $10^{9.5-10.5}$\msun, illustrated by red boxes in the left panels. In both observed and simulated systems, satellites that fell into the clusters longer ago exhibit lower fraction of stars in cold orbits.
        }
    \label{fcold_vs_mass_g2}
\end{center}
\end{figure*}

As shown in Fig.~\ref{fcold_vs_mass_g2}, in both the Fornax cluster and a TNG50 analog (halo with Group ID 2), the cold-disk fractions of recent and intermediate infallers change as a function of galaxy stellar mass, by peaking at $M_{\ast} \sim 10^{9.5}-10^{10.5}$\msun and decreasing toward the lower-mass and higher-mass ends.
 For galaxies at $\log_{10}M_{*}/\mathrm{M_\odot} = 9.5 - 10.5$, the average cold-disk fraction of recent and intermediate infallers read 0.20 (0.22) within $R_{\rm e}$($2R_{\rm e}$) for the Fornax cluster, comparable to 0.16 (0.21) for the TNG50 system. 
 On the other hand, galaxies with very early infall times (ancient infallers) have low fractions of cold disk across all studied masses. The average luminosity fraction of the dynamically cold disks within $R_{\rm e}(2R_{\rm e})$ of ancient infallers is 0.10 (0.13) in the Fornax cluster and 0.07 (0.08) in the TNG50 analog. 

 Here we used the luminosity-weighted cold-disk fraction from TNG galaxies to directly compare with observations. The luminosity-weighted cold-disk fraction is systematically higher than the mass-weighted values because of the younger stars in the disks. We include the luminosity-weighted cold-disk fraction for all galaxies from group 0-13 in Fig.~\ref{fig:fcold_light_weighted_vs_mass_all}, which can be directly compared with the mass-weighted values shown in the bottom panel of Fig.~\ref{fcold_vs_mass_all}.  
 
 Given the non-negligible galaxy-to-galaxy variations, the cold disk fractions of galaxies in TNG50 are generally consistent with those in the Fornax cluster for both recent and intermediate (combined) and ancient infallers, respectively.

 In general, the comparisons in this section lend credibility to the outcome of the TNG50 model. The dependences of the properties of satellites' cold disks, for example the stellar age and luminosity fraction, on their infall time generally agree between TNG50 (halo with Group ID 2) and the Fornax cluster -- the agreement is good but tentative.
 Given the large host-to-host and galaxy-to-galaxy variations, the agreement shown above indicates that the physical processes implemented and emerging in TNG50 lead to an evolution of cold disks in high-density environments that is compatible with observations at least in the case of one observed and one simulated cluster. Whether the model is consistent with observations of a larger statistically significant sample of clusters remains to be determined.
 
 Moreover, we deliberately did not compare the galaxy populations in the two clusters in further detail or more quantitatively. Firstly, our galaxy sample of the Fornax cluster is not complete. Secondly, the galaxies' properties may depend on the assembly history of their cluster host, and we cannot guarantee that such an assembly is exactly the same for a specific halo of TNG50 and for Fornax.

\section{Summary}\label{sec:conclusions}
In this paper we have studied the effects of cluster environment on the dynamically defined stellar structures of galaxies using the TNG50 cosmological magnetohydrodynamic simulation.

We selected 490 galaxies at $z=0$ with stellar masses in the $10^{8-12}$\msun range that reside in the most massive groups and clusters of the simulated volume at the current epoch, specifically halos with IDs 0 through 13 and with total masses of $10^{13.4-14.3}$\msun, which is similar to mass of the Fornax cluster.

We characterized each galaxy's structure by decomposing it into a dynamically cold disk and a dynamically hot non-disk component, with an approach that is very similar to what can be done with IFU observations. We defined a galaxy's infall time into its current cluster  ($t_{\rm infall}$) as the first time it crosses the virial radius of the host halo, and we traced the evolution of its star-forming gas distribution and stellar cold disk fraction back in time.

For the TNG50 cluster galaxies, we find that:
\begin{itemize}
    \item After galaxies fall into a cluster, their star-forming gas is removed and the spatial extent of their gas shrinks. We quantitatively defined the time when the star-forming gas mass starts to drop ($t_{\rm drop}$) and the transition time when the gaseous disk radius changes from increasing to decreasing ($t_{\rm trans}$) for all galaxies.\ We find that both correlate with the time the galaxies have spent into the cluster, that is to say, the time since infall ($t_{\rm infall}$). 

    \item The drop in the star-forming gas mass is rapid after infall. One billion years after infall, we find that 42\%, 76\%, and 93\% of galaxies have already lost at least 80\%, 50\%, and 20\% of their
    star-forming gas mass compared with that at $t_{\rm infall}$. In fact, 4 Gyr after $t_{\rm infall}$, 87\% of galaxies have lost at least 80\% of their star-forming gas.
    
    \item The decrease in the extent of the star-forming region is relatively rapid and pertains to all cluster galaxies. The star-forming gas extent is reduced to 20\% of its original size for 1\%, 5\%, and 25\% of satellites at 1,  2, and 4 Gyr after infall, respectively. 
    
    \item As a result of the gas mass removal and shrinkage, SF in the dynamically cold disks sharply decreases: this leads to a correlation between the stellar age of the dynamically cold disks, $t_{\rm cold}$, and the infall time, $t_{\rm infall}$ . Ancient infallers have older disks, while recent infallers have younger disks. 
    
    \item Together with tidal effects, the phenomenology above also leads to an overall reduction in the mass or light fraction of the dynamically cold stellar components in cluster galaxies. The mass fraction of the dynamically cold disk varies as a function of the galaxy stellar mass: it peaks at $M_{\ast}/M_{\odot} \sim 10^{9.5-10.5}$ and is systematically lower for galaxies that fell into a cluster earlier, irrespective of galaxy mass. 
\end{itemize}

Based on this and previous analyses of TNG50, we interpret the findings above as follows. Firstly, the removal of star-forming gas stops the possible growth of the cold disk, and it does so earlier for ancient infallers.
Furthermore, the cold disks are disrupted by tidal processes after infall, which take a long time to affect the disks in the inner $1-2R_\mathrm{e}$ of the galaxies. The ultimate effects are strongest for ancient infallers: their cold disks, even within $R_\mathrm{e}$, are almost completely disrupted by the present day. 

By comparing the galaxies in the Fornax cluster from observations to those of an analog system from TNG50, we qualitatively confirm that the galaxy formation model and the environmental effects that naturally emerge within the simulation are overall consistent with reality. In particular, we find that:
\begin{itemize}
\item In both the TNG50 Fornax-like and the Fornax clusters, the stellar ages of the cold and hot components of satellites are similar for ancient infallers, whereas the cold disks can be significantly younger than the hot components for recent infallers. In the Fornax cluster, the average stellar age difference between the dynamically cold disk and the dynamically hot non-disk component is $0 \pm 0.64$ Gyr, $1.3 \pm 1.4$ Gyr, and $1.7\pm1.7$ Gyr for ancient, intermediate, and recent infallers, respectively, while in the TNG50 cluster with Group ID 2, they are $0.2 \pm 0.2$ Gyr, $0.4 \pm 0.7$ Gyr, and $1.8 \pm 1.5$ Gyr, respectively. Such numbers quantitatively agree with each other, and the dynamically cold disk is about $\sim 2$ Gyr younger than the hot bulge defined in the same regions in recent infallers, which is consistent with galaxies in the field \citep{jin2023}. 

\item The cold-disk fractions in the galaxies of the TNG50 Fornax analog agree with those of Fornax, both in terms of values and of trends with the galaxy's stellar mass and infall time. Recent and intermediate infallers (combined) exhibit the largest fractions of cold disks in the $\log_{10} M_*/M = 9.5-10.5$ mass range: on average, 0.20 (0.22) within $R_{\rm e}$(2$R_{\rm e}$) for Fornax galaxies and 0.16 (0.21) for TNG50 analog galaxies. In both observed and simulated systems, ancient infallers have low fractions of light (i.e., mass) in cold disks across the studied satellite mass range: approximately 10\% or less.
\end{itemize}

We have demonstrated the power of the dynamical decomposition of galaxies' stellar structures. Applied to both observed and simulated systems, it has the potential to facilitate direct and quantitative comparisons. On the one hand, observational results can help in benchmarking galaxy formation models; on the other hand, taking the consistency between simulations and observations  into account, they can facilitate comprehensive physical interpretations of observational results. In an upcoming paper, we will perform a similar analysis to focus on stellar structure formation and evolution in field galaxies.

\begin{acknowledgement}
This work is supported by the Max Planck Partner's group led by LZ and AP.
LZ acknowledges the support from the National Natural Science Foundation of China under grant No. Y945271001 and the CAS Project for Young Scientists in Basic Research, Grant No. YSBR-062. 
TNG50 was realised with compute time granted by the Gauss Centre for Super-computing (GCS), under the GCS Large-Scale Project GCS-DWAR (2016; PIs Nelson/Pillepich).
EI acknowledges the support from the PRIN MUR 2022 CUP:C53D23000850006 and the INAF GO grant 2022.
EMC acknowledges the support from the MIUR grant PRIN 2017 20173ML3WW-001, INAF grant PRIN 2022 C53D23000850006, and Padua University grants DOR 2021-2023.
F.P. acknowledges support from the Agencia Estatal de Investigación del Ministerio de Ciencia e Innovación (MCIN/AEI/ 10.13039/501100011033) under grant (PID2021-128131NB-I00) and the European Regional Development Fund (ERDF) "A way of making europe".
\end{acknowledgement}

\bibliographystyle{aa}
\typeout{} 
\bibliography{main}

\begin{appendix}
\section{The luminosity fraction of the dynamically cold stellar disk}\label{app:light-weighted}
\begin{figure*}[h!]
\centering
    \includegraphics[width=1.8\columnwidth]{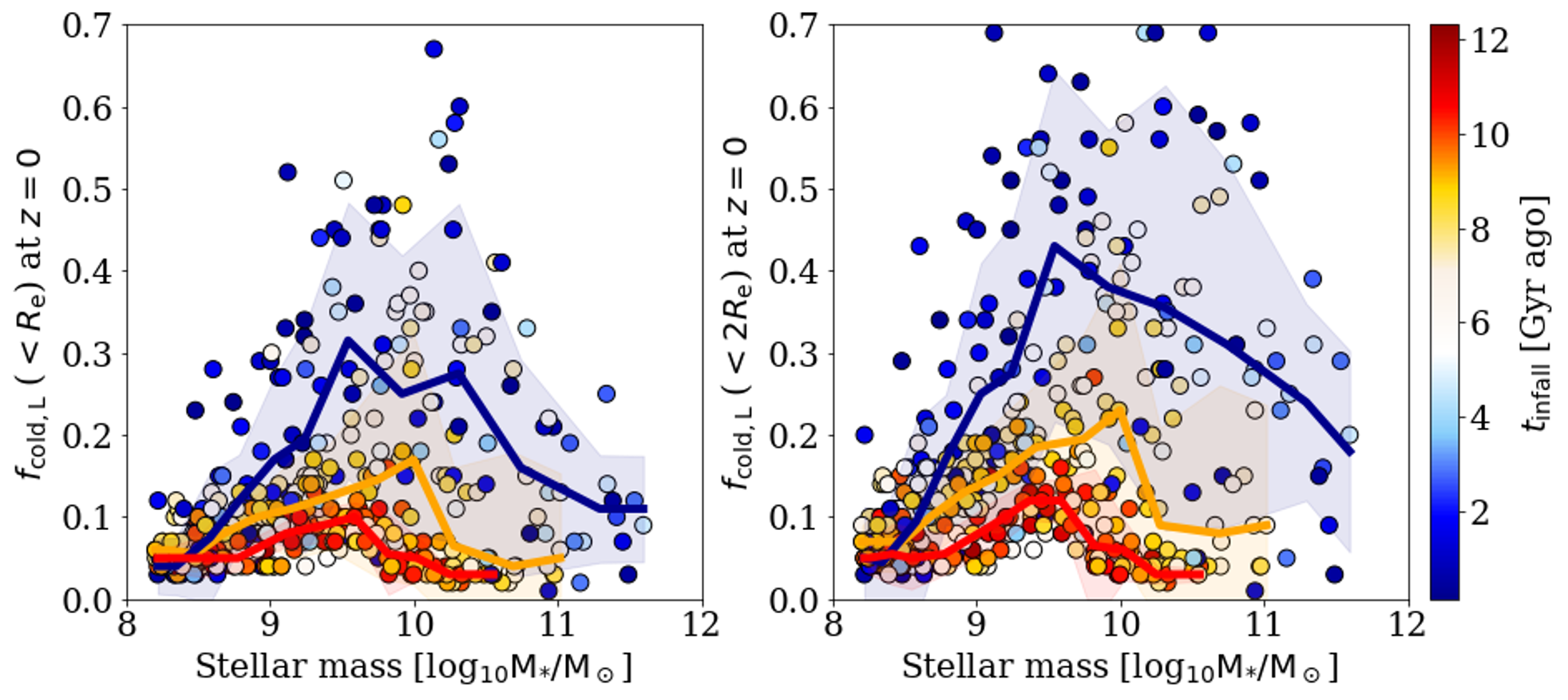}
    \caption{
        {Luminosity fraction in the dynamically cold disks in cluster galaxies according to TNG50 at the present day.} We show the luminosity-weighted cold-disk fraction of galaxies within $1R_{\rm e}$ (left) and within $2R_{\rm e}$ (right), whereby the effective radius is obtained at $z=0$. We color code the markers by the time since infall, according to the color bars. Red, yellow, and blue curves and shaded regions denote population medians and 1$\sigma$ variations of the stellar cold-disk fraction in bins of present-day galaxy stellar mass for ancient, intermediate, and recent infallers, respectively. }
    \label{fig:fcold_light_weighted_vs_mass_all}
\end{figure*}
\end{appendix}

\end{document}